\documentclass[12pt,preprint,aps,prd,nofootinbib,superscriptaddress, tightenlines]{revtex4}

\usepackage[OT1]{fontenc}
\usepackage[utf8]{inputenc}
\usepackage{epsfig}
\usepackage{epstopdf}
\usepackage[dvipsnames]{xcolor}
\usepackage{graphics}
\usepackage{lscape}
\usepackage{rotfloat}
\usepackage{rotating}
\usepackage{amsmath}
\usepackage{amssymb}
\usepackage{graphicx}
\usepackage{flexisym}
\usepackage{breqn}
\usepackage{mathtools}
\usepackage{slashed}
\usepackage[hidelinks=true]{hyperref}
\usepackage{enumitem}

\newcommand{\bdm}{\begin{dmath}}
\newcommand{\edm}{\end{dmath}}
\newcommand{\bdms}{\begin{dmath*}}
\newcommand{\edms}{\end{dmath*}}
\newcommand{\bdg}{\begin{dgroup*}}
\newcommand{\edg}{\end{dgroup*}}  
\newcommand{\be}{\begin{equation}}
\newcommand{\ee}{\end{equation}}
\newcommand{\bea}{\begin{eqnarray}} 
\newcommand{\eea}{\end{eqnarray}}
\def\openone{\leavevmode\hbox{\small1\kern-3.3pt\normalsize1}}

\newcommand{\MSbar}{\overline{\textrm{MS}}} 
\newcommand{\logq}{\ln \left( \frac{q^2}{\bar{\mu}^2} \right)}
\newcommand{\logqsq}{\ln^2 \left( \frac{q^2}{\bar{\mu}^2} \right)}
\newcommand{\logqb}{\ln \left( \frac{\bar{q}^2}{\bar{\mu}^2} \right)}
\newcommand{\logqbsq}{\ln^2 \left( \frac{\bar{q}^2}{\bar{\mu}^2} \right)}

\definecolor{mygreen}{RGB}{0,153,0}

\definecolor{myorange}{RGB}{255,130,0}

\newcommand{\cev}[1]{\reflectbox{\ensuremath{\vec{\reflectbox{\ensuremath{#1}}}}}}

\begin{document}
\title{Two-loop renormalization and mixing of gluon and quark energy-momentum tensor operators}
\author{George Panagopoulos}
\email{gpanago@stanford.edu}
\affiliation{$\it{Department \ of \ Physics, \ Stanford \ University, \ California \ 943052004, \ USA}$ \vspace{0.1cm}}
\author{Haralambos Panagopoulos} 
\email{haris@ucy.ac.cy}
\affiliation{$\it{Department \ of \ Physics, \ University \ of \ Cyprus, \ POB \ 20537, \ 1678, \ Nicosia, \ Cyprus}$ \vspace{0.5cm}}
\author{Gregoris Spanoudes \vspace{0.5cm}} 
\email{spanoudes.gregoris@ucy.ac.cy}
\affiliation{$\it{Department \ of \ Physics, \ University \ of \ Cyprus, \ POB \ 20537, \ 1678, \ Nicosia, \ Cyprus}$ \vspace{0.5cm}}
\begin{abstract}
\vspace{0.5cm} In this paper, we present one- and two-loop results for the renormalization of the gluon and quark gauge-invariant operators which appear in the definition of the QCD energy-momentum tensor, in dimensional regularization. To this end, we consider a variety of Green's functions with different incoming momenta. We identify the set of twist-2 symmetric traceless and flavor singlet operators which mix among themselves and we calculate the corresponding mixing coefficients for the nondiagonal components. We also provide results for some appropriate regularization-independent (RI$'$)-like schemes, which address this mixing, and we discuss their application to nonperturbative studies via lattice simulations. Finally, we extract the one- and two-loop expressions of the conversion factors between the proposed RI$'$ and the $\MSbar$ schemes. From our results regarding the $\MSbar$-renormalized Green's functions, one can easily derive conversion factors relating numerous variants of RI$'$-like schemes to $\MSbar$.

To make our results easily accessible, we also provide them as Supplemental Material, in the form of a {\it Mathematica} input file and, also, an equivalent text file.     
\end{abstract}
\hspace{-5mm}\begin{minipage}{\textwidth}
\maketitle
\end{minipage}
\renewcommand{\thefootnote}{\alph{footnote}}
\footnotetext{Electronic addresses: ${}^{a}$ gpanago@stanford.edu, \ ${}^{b}$ haris@ucy.ac.cy, \ ${}^{c}$ spanoudes.gregoris@ucy.ac.cy}
\renewcommand{\thefootnote}{\arabic{footnote}}
\section{Introduction}
\label{sec:intro}

An important open question in Hadronic Physics is the hadron spin decomposition, i.e., the distribution of hadron spin among its constituent particles. It is well known, by recent experiments, that contributions to the hadron spin arise not only from valence quarks, but also from polarized gluons, as well as sea quarks. Therefore, it is understood that the complete picture of the spin content of a hadron requires taking into account its nonperturbative nature, including gluon and quark disconnected contributions. Useful quantities which give important input to the study of hadron spin structure are the quark and gluon average momentum fractions \cite{Ji:1998pc}. Their nonperturbative determination in nucleons is currently under investigation by a number of research groups \cite{Yang:2018nqn, Alexandrou:2020sml, Mondal:2020cmt}, and so far, the outcomes are very promising for the correct extraction of the nucleon spin decomposition. However, there are still many challenges that need to be faced, including the complete renormalization of these quantities. 

\bigskip
  
Recent progress in simulating QCD on the lattice has allowed the first \textit{ab initio} studies of more demanding quantities in hadron structure, involving, e.g., gluon and quark flavor singlet operators; these quantities suffer from two issues: the increased statistical noise and the presence of mixing with other operators. Vigorous efforts in addressing the former include optimized algorithms and increased statistics. The latter issue has additional difficulties: the operators which mix among themselves are typically defined in perturbation theory and may involve gauge-variant (GV) terms and ghost fields; thus, their nonperturbative calculation, by compact lattice simulations, is not feasible. There remains still a number of conceptual questions to be resolved before a viable nonperturbative evaluation of mixing effects can be implemented. Studying the mixing pattern in higher orders of perturbation theory can give important guidance for the corresponding elimination of operator mixing nonperturbatively.   

\bigskip

In this work, we study the renormalization and mixing of gluon and quark singlet gauge-invariant operators appearing in the definition of the QCD energy-momentum tensor (EMT). These operators are employed in the calculation of the quark and gluon average momentum fractions in hadrons. In terms of the gluon field $A_\mu^a$ and quark field $\psi_f$, they are defined as \cite{Caracciolo:1991cp}\footnote{We will refer to ${O_1}_{\mu \nu}$ and ${O_2}_{\mu \nu}$ as the ``gluon'' and ``quark'' EMT operators, respectively.}
\begin{eqnarray}
{O_1}_{\mu \nu} &=& F^{a}_{\mu \rho} F^{a}_{\nu \rho} - \frac{1}{d} \delta_{\mu \nu} F^{a}_{\rho \sigma} F^{a}_{\rho \sigma}, \label{O1} \\
{O_2}_{\mu \nu} &=& \sum_{f} \left[ \frac{1}{2} \left( \bar{\psi}_f \gamma_\mu \overleftrightarrow{D}_\nu \psi_f + \bar{\psi}_f \gamma_\nu \overleftrightarrow{D}_\mu \psi_f \right) - \frac{1}{d} \delta_{\mu \nu} \left( \bar{\psi}_f \gamma_\rho \overleftrightarrow{D}_\rho \psi_f \right) \right], \label{O2}
\end{eqnarray}

\smallskip

\hspace{-0.5cm}$\Big[F^a_{\mu \nu} \equiv \partial_\mu A^a_\nu - \partial_\nu A^a_\mu - g f^{abc} A^b_\mu A^c_\nu$ is the field strength tensor and $f^{abc}$ are the $SU(N_c)$ structure constants; $\overleftrightarrow{D}_\mu \equiv (1 / 2) ( \overrightarrow{D}_\mu - \overleftarrow{D}_\mu )$ is the symmetrized covariant derivative, and $\overrightarrow{D}_\mu \equiv \overrightarrow{\partial}_\mu + i g A_\mu$, $\overleftarrow{D}_\mu \equiv \overleftarrow{\partial}_\mu - i g A_\mu$ are the left and right covariant derivatives, respectively. The index $f$ in Eq. \eqref{O2} is summed over $N_f$ quark flavors, and $d$ is the number of Euclidean space-time dimensions. Greek indices ($\mu, \nu, \rho, \sigma$) and Latin indices ($a, b, c$) refer to the Lorentz and SU($N_c$) groups, respectively. A summation over repeated indices is implied.$\Big]$

\bigskip

Given that we will consider mass-independent renormalization schemes, a mass term has been omitted from the definition of $O_{2 \mu \nu}$. These operators can mix with GV operators, which vanish when inserted in matrix elements between physical states. However, the mixing with GV  operators cannot be neglected: the standard perturbative procedure for the correct extraction of the mixing coefficients entails calculating bare Green's functions (GFs) of GV operators with elementary external fields. The goal of our study is twofold as follows: 
\begin{enumerate}
\item To identify the set of twist-2 symmetric operators which mix with the gluon and quark EMT operators, and to provide an appropriate regularization-independent (RI$'$)-like scheme, which correctly addresses this mixing. 
\item To compute the conversion factors from the proposed RI$'$-like schemes to the $\MSbar$ scheme. 
\end{enumerate}
We calculate a total of ten one-particle-irreducible (1PI) amputated Green's functions with operator insertions up to two loops in dimensional regularization (DR). In order to be able to extract the mixing coefficients in an unambiguous way, we consider Green's functions with different incoming momenta. 

\bigskip

The renormalization factors of gluon and quark EMT operators can be extracted by studying either the diagonal or the nondiagonal components of the operators. As the EMT operators are traceless, it becomes difficult to disentangle the signal of the diagonal part in lattice simulations from the corresponding pure trace. The mixing pattern of the nondiagonal components is simpler comparing to the diagonal ones. For this reason, we choose to consider only nondiagonal components. 

\bigskip 

We investigated possible ways of defining an appropriate RI$'$-type scheme, which can be applied in the nonperturbative studies on the lattice. Green's functions of GV operators are difficult to obtain nonperturbatively on the lattice due to a number of obstacles: GV operators [Becchi-Rouet-Stora-Tyutin (BRST) variations and operators which vanish by the equations of motion (EOM)] are defined in a perturbative manner, including gauge-fixing terms, which are not well defined in the Landau gauge (they contain terms proportional to $1/\alpha$, where $\alpha$ is the gauge-fixing parameter and $\alpha = 0$ in the Landau gauge) and ghost fields. Such terms cannot be studied by compact lattice simulations. In our study, we discuss some possible approaches to overcome this issue.

\bigskip
   
A novel aspect of this calculation is the extraction of the mixing matrix to two-loop order. A number of previous perturbative and nonperturbative studies of EMT have been carried out in both continuum and lattice regularizations. A one-loop calculation of the mixing matrix in the continuum is presented in Ref. \cite{Collins:1994ee}. Earlier studies of flavor singlet operator renormalization in the continuum can be found in Refs.~\cite{Dixon:1974ss,Matiounine:1998ky}. Corresponding one-loop calculations on the lattice are considered in Refs. \cite{Capitani:1994qn, Alexandrou:2016ekb, Yang:2016xsb}. A conserved EMT for lattice gauge theories is constructed in Refs. \cite{Caracciolo:1989pt, Caracciolo:1991cp} to one-loop level. Recent nonperturbative studies of the renormalization of gluon and quark EMT operators have been performed in lattice QCD simulations \cite{Yang:2018bft, Yang:2018nqn, Shanahan:2018pib, Alexandrou:2020sml, Mondal:2020cmt}. A promising investigation for determining a conserved EMT nonperturbatively on the lattice in a regularization group invariant (RGI) scheme is given in Refs. \cite{Giusti:2015daa, DallaBrida:2020gux}.

\bigskip   
	
The outline of this paper is as follows: in Sec. \ref{TheoreticalAnalysis}, we provide a theoretical analysis of the renormalization of gluon and quark EMT operators based on the Joglekar-Lee theorems and Ward identities (WIs) of BRST-invariant operators and of conserved quantities. Section \ref{Calculation Setup} contains the calculation setup including details on the calculated Green's functions, description of the proposed renormalization schemes, and the conversion to the $\MSbar$ scheme. Our main results are presented in Sec. \ref{Results} for the $\MSbar$-renormalized Green's functions, the renormalization functions, and the conversion factors between the RI$'$ and the $\MSbar$ schemes. In Sec. \ref{Nonperturbative renormalization}, we discuss the application of the proposed RI$'$ schemes in the nonperturbative studies on the lattice, while in Sec. \ref{Summary} we conclude.      

\section{Theoretical Analysis}
\label{TheoreticalAnalysis}

According to the Joglekar-Lee theorems\cite{Joglekar:1975nu}, a gauge-invariant operator $\mathcal{O}$ can mix with three classes of  operators which have the same transformations under global symmetries (e.g., Lorentz, or hypercubic on the lattice, global SU($N_c$) transformations, etc.) and whose dimension is lower or equal to that of $\mathcal{O}$:
\begin{enumerate}
\item Class G: Gauge-invariant operators
\item Class A: BRST variations of some operator
\item Class B: Operators which vanish by the EOM
\end{enumerate}
Any other operators which respect the same global symmetries, but do not belong to the above classes, can at most have finite mixing with $\mathcal{O}$ \cite{Joglekar:1975nu}. In this respect and given that gluon and quark EMT operators are two-index traceless symmetric of dimension 4, the full set of twist-2 operators which mix among themselves, compatibly with Euclidean rotational symmetry, is the following \cite{Caracciolo:1991cp}: 
\begin{eqnarray}
{O_1}_{\mu \nu} &=& F^{a}_{\mu \rho} F^{a}_{\nu \rho} - \frac{1}{d} \delta_{\mu \nu} F^{a}_{\rho \sigma} F^{a}_{\rho \sigma}, \label{Op1} \\
\nonumber \\
{O_2}_{\mu \nu} &=& \sum_{f} \left[ \frac{1}{2} \left( \bar{\psi}_f \gamma_\mu \overleftrightarrow{D}_\nu \psi_f + \bar{\psi}_f \gamma_\nu \overleftrightarrow{D}_\mu \psi_f \right) - \frac{1}{d} \delta_{\mu \nu} \left( \bar{\psi}_f \gamma_\rho \overleftrightarrow{D}_\rho \psi_f \right) \right], \label{Op2} \\
\nonumber \\
{O_3}_{\mu \nu} &=& \frac{1}{\alpha} \left[ \left( \partial_\mu A_\nu^a + \partial_\nu A_\mu^a \right) \partial_\rho A_\rho^a - \frac{2}{d} \delta_{\mu \nu} \partial_\rho A_\rho^a \partial_\sigma A_\sigma^a \right] \nonumber \\
&& - \left[ \bar{c}^a \partial_\mu {\left( D_\nu c \right)}^a + \bar{c}^a \partial_\nu {\left( D_\mu c \right)}^a - \frac{2}{d} \delta_{\mu \nu} \bar{c}^a \partial_\rho {\left( D_\rho c \right)}^a \right], \label{Op3} \\
\nonumber \\
{O_4}_{\mu \nu} &=& - \frac{1}{\alpha} \left[ \left( A_\mu^a \partial_\nu + A_\nu^a \partial_\mu \right) \left( \partial_\rho A_\rho^a \right) - \frac{2}{d} \delta_{\mu \nu} A_\rho^a \partial_\rho \partial_\sigma A_\sigma^a \right] \nonumber \\
&& \quad + \left[ \partial_\mu \bar{c}^a {D_\nu c}^a + \partial_\nu \bar{c}^a {D_\mu c}^a - \frac{2}{d} \delta_{\mu \nu} \partial_\rho \bar{c}^a {D_\rho c}^a \right], \label{Op4} \\
\nonumber \\
{O_5}_{\mu \nu} &=& A_\mu^a \frac{\delta S}{\delta A_\nu^a} + A_\nu^a \frac{\delta S}{\delta A_\mu^a} - \frac{2}{d} \delta_{\mu \nu} A_\rho^a \frac{\delta S}{\delta A_\rho^a}, \label{Op5}
\end{eqnarray}
where a summation over repeated indices is implied; $c^a$ and $\bar{c}^a$ are the ghost and antighost fields, respectively, and $S$ is the QCD action,
\begin{equation}
S = \int d^d x \left[ \frac{1}{4} F^a_{\rho \sigma} F^a_{\rho \sigma} + \bar{\psi} \gamma_\rho D_\rho \psi + \frac{1}{2 \alpha} \partial_\rho A^a_\rho \partial_\sigma A^a_\sigma - \bar{c}^a \partial_\rho {\left( D_\rho c \right)}^a \right].
\end{equation}
${O_1}_{\mu \nu}$ and ${O_2}_{\mu \nu}$ are class G operators, ${O_3}_{\mu \nu}$ and ${O_4}_{\mu \nu}$ belong to class A, and ${O_5}_{\mu \nu}$ is a class B operator.

\bigskip

In the absence of quarks, operators ${O_3}_{\mu \nu}$, ${O_4}_{\mu \nu}$, ${O_5}_{\mu \nu}$ are the only operators that can mix with ${O_1}_{\mu \nu}$~\cite{Caracciolo:1991cp}. Upon introducing quarks (and, therefore, also ${O_2}_{\mu \nu}$), one must investigate whether any additional operators can mix. The answer is negative as follows:
\begin{itemize}
\item The only other class G operator which is traceless and symmetric,
  \begin{equation}
    \sum_{f} \left\{ \frac{1}{4} \left[ \partial_\mu \left(\bar{\psi}_f \gamma_\nu \psi_f \right) + \partial_\nu \left( \bar{\psi}_f \gamma_\mu \psi_f \right) \right]- \frac{1}{2 d} \delta_{\mu \nu} \partial_\rho \left( \bar{\psi}_f \gamma_\rho \psi_f \right) \right\}, 
  \end{equation}
being odd under charge conjugation, is excluded. By the same token, operator ${O_2}_{\mu \nu}$ with the symmetrized covariant derivatives replaced by only right or left covariant derivatives is not considered since it is not a pure eigenstate of charge conjugation.
\item There are no operator candidates containing fermions in class A, as any two-index operator with fermion and antifermion fields will lead, under BRST transformations, to an operator of dimension at least 5.
\item The only potential class B operator stemming from the fermion EOM is pure trace, and thus it is excluded.
\end{itemize}

\bigskip

On the lattice, where Lorentz symmetry is replaced by hypercubic symmetry, diagonal ($\mu = \nu$) and nondiagonal ($\mu \neq \nu$) components of traceless symmetric operators belong to different representations of the hypercubic group, and thus, they renormalize differently. As we are interested in constructing a renormalization scheme applicable to the lattice, we must renormalize diagonal and nondiagonal components separately. However, their corresponding renormalized Green's functions will be components of a common multiplet in the continuum limit, as it happens in continuum regularizations. In this study, we focus on the renormalization of the nondiagonal components of the EMT operators, because they give more accurate results in lattice simulations when inserted in matrix elements between physical states \cite{Alexandrou:2020sml}. From now on, when we refer to ${O_i}_{\mu \nu}$, $(i = 1, 2, \ldots, 5)$, it is meant that $\mu \neq \nu$.    

\bigskip

Operators ${O_1}_{\mu \nu}, {O_2}_{\mu \nu}, \ldots, {O_5}_{\mu \nu}$ have some interesting properties which give us an important input in the study of their renormalization. Let us define the mixing matrix $Z$ as follows:
\begin{equation}
O_i^Y = \sum_{j=1}^5 Z^{Y,X}_{ij} O_j^X, \ (i = 1,2,3,4,5), \label{Z_O}
\end{equation}
where $O_i^X (O_i^Y)$ is the bare (renormalized) operator ${O_i}_{\mu \nu}$ in the X regularization (Y renormalization) scheme. Here, to simplify the notation we omit the Lorentz indices $\mu$, $\nu$. The sum $O_1^{\rm DR} + O_2^{\rm DR} + O_4^{\rm DR}$ in dimensional regularization gives the nondiagonal Belinfante symmetrized EMT \cite{BELINFANTE1940449}, which is a conserved quantity.\footnote{Note that a possible variant of a symmetrized EMT includes an admixture of the operator $O_5^{\rm DR}$, which is also a conserved quantity.} As a consequence, this combination of operators has zero anomalous dimension and thus, it is finite. This is also true for the class B operator $O_5^{\rm DR}$. This means that, in the $\overline{\rm MS}$ scheme, we have
\begin{equation}
O_1^{\overline{\rm MS}} + O_2^{\overline{\rm MS}} + O_4^{\overline{\rm MS}} = O_1^{\rm DR} + O_2^{\rm DR} + O_4^{\rm DR}, \label{finite1}
\end{equation}
\begin{equation}
O_5^{\overline{\rm MS}} = O_5^{\rm DR}. \label{finite2}
\end{equation}
Replacing Eq. \eqref{Z_O} into Eqs. (\ref{finite1}, \ref{finite2}), the following relations between the elements of the mixing matrix are extracted \cite{Ji:1995sv}:  
\begin{eqnarray}
Z_{11}^{\overline{\rm MS},{\rm DR}} + Z_{21}^{\overline{\rm MS},{\rm DR}} + Z_{41}^{\overline{\rm MS},{\rm DR}} &=& 1, \label{cons1}\\
Z_{12}^{\overline{\rm MS},{\rm DR}} + Z_{22}^{\overline{\rm MS},{\rm DR}} + Z_{42}^{\overline{\rm MS},{\rm DR}} &=& 1, \label{cons2}\\
Z_{13}^{\overline{\rm MS},{\rm DR}} + Z_{23}^{\overline{\rm MS},{\rm DR}} + Z_{43}^{\overline{\rm MS},{\rm DR}} &=& 0, \label{cons3}\\
Z_{14}^{\overline{\rm MS},{\rm DR}} + Z_{24}^{\overline{\rm MS},{\rm DR}} + Z_{44}^{\overline{\rm MS},{\rm DR}} &=& 1, \label{cons4}\\
Z_{15}^{\overline{\rm MS},{\rm DR}} + Z_{25}^{\overline{\rm MS},{\rm DR}} + Z_{45}^{\overline{\rm MS},{\rm DR}} &=& 0, \label{cons5}\\
Z_{51}^{\overline{\rm MS},{\rm DR}} = Z_{52}^{\overline{\rm MS},{\rm DR}} = Z_{53}^{\overline{\rm MS},{\rm DR}} = Z_{54}^{\overline{\rm MS},{\rm DR}} &=& 0, \label{cons6} \\
Z_{55}^{\overline{\rm MS},{\rm DR}} &=& 1. \label{cons7}  
\end{eqnarray}

\bigskip

Furthermore, according to the Joglekar-Lee theorems\cite{Joglekar:1975nu}, the mixing matrix (at least) in DR and the $\overline{\rm MS}$ scheme is block triangular, i.e., class A operators cannot mix with class G operators, and class B operators cannot mix with class G and class A operators; thus,  
\begin{eqnarray}
Z_{31}^{\overline{\rm MS},{\rm DR}} = Z_{32}^{\overline{\rm MS},{\rm DR}} = Z_{41}^{\overline{\rm MS},{\rm DR}} = Z_{42}^{\overline{\rm MS},{\rm DR}} &=& 0, \\
Z_{51}^{\overline{\rm MS},{\rm DR}} = Z_{52}^{\overline{\rm MS},{\rm DR}} = Z_{53}^{\overline{\rm MS},{\rm DR}} = Z_{54}^{\overline{\rm MS},{\rm DR}} &=& 0. \label{triag}
\end{eqnarray}

\bigskip

Additional relations between the elements of the mixing matrix can be extracted by studying WIs which contain operators $O_i$. Let us consider the following WI:
\begin{equation}
\delta_{\rm BRST} \langle \partial_\rho A_\rho^a (x) \ O_i^X (y) \ \bar{c}^b (z) \rangle = 0,
\label{WI_BRST}
\end{equation} 
where $\delta_{\rm BRST}$ is the BRST operator. Because of the BRST invariance of both action and class G, A, and B operators (modulo equations of motion), Eq. \eqref{WI_BRST} takes the following form:
\begin{eqnarray}
&& \frac{1}{\alpha} \langle \partial_\rho A_\rho^a (x) \ O_1^X (y) \ \partial_\sigma A_\sigma^b (z) \rangle = 0, \label{WI_BRST2a} \\
&& \frac{1}{\alpha} \langle \partial_\rho A_\rho^a (x) \ O_2^X (y) \ \partial_\sigma A_\sigma^b (z) \rangle = 0, \label{WI_BRST2b} \\
&& \frac{1}{\alpha} \langle \partial_\rho A_\rho^a (x) \ \left( O_4^X (y) - O_5^X (y) \right) \ \partial_\sigma A_\sigma^b (z) \rangle = 0. \label{WI_BRST2c}
\end{eqnarray}
In momentum space, they read
\begin{eqnarray}
\alpha \ p_\rho q_\sigma {\langle A_\rho^a (p) \ O_1^X (-p-q) \ A_\sigma^b (q) \rangle}_{\rm amp} &=& 0, \quad \forall \ p, q, \alpha, \label{WI_BRST3a} \\
\alpha \ p_\rho q_\sigma {\langle A_\rho^a (p) \ O_2^X (-p-q) \ A_\sigma^b (q) \rangle}_{\rm amp} &=& 0, \quad \forall \ p, q, \alpha, \label{WI_BRST3b} \\
\alpha \ p_\rho q_\sigma {\langle A_\rho^a (p) \ \left(O_4^X (-p-q) - O_5^X (-p-q) \right) \ A_\sigma^b (q) \rangle}_{\rm amp} &=& 0, \quad \forall \ p, q, \alpha. \label{WI_BRST3c}
\end{eqnarray}
By replacing the bare operators with the renormalized ones, the above relations also hold (at least) in the $\MSbar$ scheme. This is proved by the following arguments. Let us consider, e.g., the Green's function of operator $O_1$ in the $Y$ renormalization scheme: $\alpha \ p_\rho q_\sigma {\langle A_\rho^a (p) \ O_1^Y (-p-q) \ A_\sigma^b (q) \rangle}_{\rm amp}$. Using Eqs. (\ref{Z_O}, \ref{WI_BRST3a}, \ref{WI_BRST3b}, \ref{WI_BRST3c}), the Green's function takes the following form:
\begin{eqnarray}
&& \qquad \qquad Z^{Y,X}_{13} \left( \alpha \ p_\rho q_\sigma {\langle A_\rho^a (p) \ O_3^X (-p-q) \ A_\sigma^b (q) \rangle}_{\rm amp}\right) \nonumber \\
&+& (Z^{Y,X}_{14} + Z^{Y,X}_{15}) \left( \alpha \ p_\rho q_\sigma {\langle A_\rho^a (p) \ O_4^X (-p-q) \ A_\sigma^b (q) \rangle}_{\rm amp}\right). \label{GF_O1}
\end{eqnarray}   
Operators $O_3$ and $O_4$ differ by total derivative terms; this gives rise to different Lorentz structures in the Green's function from each operator, when $p+q \neq 0$. Thus, Eq. \eqref{GF_O1} is finite, when the poles from the $O_3$ and $O_4$ terms vanish separately, i.e., each one of the two summands in Eq. \eqref{GF_O1} must be free of poles. However, as $Z^{Y,X}_{13}$ and $Z^{Y,X}_{14} + Z^{Y,X}_{15}$ have no $\mathcal{O} (g^0)$ contributions, they must be zero to all orders in perturbation theory, at least for  $Y = \MSbar$ and $X={\rm DR}$. By similar arguments, we extract the following relations between the renormalization factors $Z^{\MSbar,{\rm DR}}_{ij}$:
\begin{eqnarray}
Z_{13}^{\MSbar,{\rm DR}} = Z_{23}^{\MSbar,{\rm DR}} &=& 0, \\
Z_{43}^{\MSbar,{\rm DR}} - Z_{53}^{\MSbar,{\rm DR}} &=& 0, \label{BRS2} \\ 
Z_{14}^{\MSbar,{\rm DR}} + Z_{15}^{\MSbar,{\rm DR}} = Z_{24}^{\MSbar,{\rm DR}} + Z_{25}^{\MSbar,{\rm DR}}  &=& 0, \\
Z_{44}^{\MSbar,{\rm DR}} + Z_{45}^{\MSbar,{\rm DR}} - Z_{54}^{\MSbar,{\rm DR}} - Z_{55}^{\MSbar,{\rm DR}} &=& 0. \label{BRS4}
\end{eqnarray}   
Combining Eqs. (\ref{BRS2}, \ref{BRS4}) with (\ref{cons6}, \ref{cons7}), we take
\begin{eqnarray}
Z_{43}^{\MSbar,{\rm DR}} &=& 0, \\
Z_{44}^{\MSbar,{\rm DR}} + Z_{45}^{\MSbar,{\rm DR}} &=& 1.
\end{eqnarray}
As we see, operators $O_1$, $O_2$, $O_4$, and $O_5$ do not mix with $O_3$ in ($\MSbar$, DR). Also, operators $O_1$ and $O_2$ mix with the combination $O_4 - O_5$. However, in a different renormalization scheme (e.g., RI$'$), these conclusions are not mandatory. 

\bigskip

Further WIs are derived for 1PI Green's functions with conserved quantities. For example, let us consider the following relation:
\begin{eqnarray}
\sum_\mu \partial_\mu \tilde{T}^{\mu \nu} & \equiv & \sum_\mu \partial_\mu \Bigg[ {O_1^{\rm DR}}_{\mu \nu} + {O_2^{\rm DR}}_{\mu \nu} + {O_4^{\rm DR}}_{\mu \nu} - {O_5^{\rm DR}}_{\mu \nu} \nonumber \\
&& \qquad \qquad + \frac{2}{d} \delta_{\mu \nu} \left( \sum_{\rho, \sigma} \partial_\rho \left( A^a_\rho \partial_\sigma A^a_{\sigma} \right) - \sum_{\rho} \partial_\rho \bar{c}^a D_\rho c^a \right) \Bigg]  \nonumber \\
& \stackrel{d \rightarrow 4}{=} & - \frac{S \cev{\delta}}{\delta c^a} \partial_\nu c^a - \partial_\nu \bar{c}^a \frac{\vec{\delta} S}{\delta \bar{c}^a} - \frac{S \cev{\delta}}{\delta \psi} \partial_\nu \psi - \partial_\nu \bar{\psi} \frac{\vec{\delta} S}{\delta \bar{\psi}} \nonumber \\
&& - \partial_\nu A^a_\mu \frac{\delta S}{\delta A^a_\mu} - \partial_\mu \left( A^a_\mu \frac{\delta S}{\delta A^a_\nu} \right) + \frac{1}{2} \partial_\nu \left( A^a_\mu \frac{\delta S}{\delta A^a_\mu} \right) \nonumber \\
&& + \frac{3}{8} \partial_\nu \left( \bar{\psi} \frac{\vec{\delta} S}{\delta \bar{\psi}} + \frac{S \cev{\delta}}{\delta \psi} \psi \right) -\frac{1}{4} \partial_\mu \left( \bar{\psi} \sigma_{\mu \nu} \frac{\vec{\delta} S}{\delta \bar{\psi}} - \frac{S \cev{\delta}}{\delta \psi} \sigma_{\mu \nu} \psi \right),
\label{WI_cons}
\end{eqnarray} 
where diagonal components of $O_i^{\rm DR}$ are also involved; $\sigma_{\mu \nu} \equiv [\gamma_{\mu}, \gamma_{\nu}] / 2$. The quantity $\tilde{T}^{\mu \nu}$ is conserved in the limit $d \rightarrow 4$. Inserting the above equation under the functional integral of the effective action $\Gamma$, a master equation is extracted which is suitable for generating WIs,
\begin{eqnarray}
\sum_\mu \partial_\mu \tilde{T}^{\mu \nu} \stackrel{d \rightarrow 4}{=} && - \frac{\Gamma \cev{\delta}}{\delta c^a} \partial_\nu c^a - \partial_\nu \bar{c}^a \frac{\vec{\delta} \Gamma}{\delta \bar{c}^a} - \frac{\Gamma \cev{\delta}}{\delta \psi} \partial_\nu \psi - \partial_\nu \bar{\psi} \frac{\vec{\delta} \Gamma}{\delta \bar{\psi}} \nonumber \\
&& - \partial_\nu A^a_\mu \frac{\delta \Gamma}{\delta A^a_\mu} - \partial_\mu \left( A^a_\mu \frac{\delta \Gamma}{\delta A^a_\nu} \right) + \frac{1}{2} \partial_\nu \left( A^a_\mu \frac{\delta \Gamma}{\delta A^a_\mu} \right) \nonumber \\
&& + \frac{3}{8} \partial_\nu \left( \bar{\psi} \frac{\vec{\delta} \Gamma}{\delta \bar{\psi}} + \frac{\Gamma \cev{\delta}}{\delta \psi} \psi \right) -\frac{1}{4} \partial_\mu \left( \bar{\psi} \sigma_{\mu \nu} \frac{\vec{\delta} \Gamma}{\delta \bar{\psi}} - \frac{\Gamma \cev{\delta}}{\delta \psi} \sigma_{\mu \nu} \psi \right). \qquad 
\label{WI_cons2}
\end{eqnarray}  
After some operations, two useful WIs are produced for zero momentum transfer which are as follows\footnote{For more details about the derivation of these WIs, we refer to \cite{Caracciolo:1989pt}.}: 
\begin{eqnarray}
&& \left \langle A^a_\rho (q) \left[ {O_1^{\rm DR}}_{\mu \nu} (0) + {O_2^{\rm DR}}_{\mu \nu} (0) + {O_4^{\rm DR}}_{\mu \nu} (0) - {O_5^{\rm DR}}_{\mu \nu} (0) \right] A^b_\sigma (-q) \right\rangle_{\rm amp} = \nonumber \\
&& \qquad \quad \ - \frac{1}{2} \left( \delta_{\mu \rho} {\left(D^{-1} (q) \right)}^{ab}_{\nu \sigma} + \delta_{\nu \rho} {\left(D^{-1} (q) \right)}^{ab}_{\mu \sigma} + \delta_{\mu \sigma} {\left(D^{-1} (q) \right)}^{ab}_{\nu \rho} + \delta_{\nu \sigma} {\left(D^{-1} (q) \right)}^{ab}_{\mu \rho} \right) \qquad \nonumber \\
&& \qquad \quad \ + \frac{1}{2} \left( q_\mu \frac{\partial}{\partial q_\nu} + q_\nu \frac{\partial}{\partial q_\mu} \right) {\left(D^{-1} (q) \right)}^{ab}_{\rho \sigma}, \qquad \label{WI_cons3a} 
\end{eqnarray}
\begin{eqnarray}
&& \left \langle \psi (q) \left[ {O_1^{\rm DR}}_{\mu \nu} (0) + {O_2^{\rm DR}}_{\mu \nu} (0) + {O_4^{\rm DR}}_{\mu \nu} (0) - {O_5^{\rm DR}}_{\mu \nu} (0) \right] \bar{\psi} (q) \right\rangle_{\rm amp} = \nonumber \\
&& \qquad \qquad \frac{1}{2} \left( q_\mu \frac{\partial}{\partial q_\nu} + q_\nu \frac{\partial}{\partial q_\mu} \right) S^{-1} (q), \label{WI_cons3b}
\end{eqnarray}
where ${\left(D^{-1} (q) \right)}^{ab}_{\mu \nu}$ and $S^{-1} (q)$ are the inverse gluon and quark propagators, respectively. Note that in Eqs. (\ref{WI_cons3a}, \ref{WI_cons3b}), indices $\mu$ and $\nu$ are taken to be different. The above relations can be useful for the construction of the nondiagonal elements of EMT on the lattice.

\section{Calculation Setup}
\label{Calculation Setup}

In this section, we briefly introduce the setup of our calculation. We provide details on the calculated Green's functions, on the renormalization prescriptions that we use in the presence of operator mixing, and on the conversion factors.   

\subsection{Green's functions}
In order to study the renormalization of the five operators defined in Eqs. \eqref{Op1} -- \eqref{Op5}, we must consider a variety of GFs with different external elementary fields and different incoming momenta. We consider a total of five GFs with external gluon fields for two different choices of incoming momenta and five GFs with external fermion fields for one choice of incoming momenta. Based on the different Lorentz and Dirac structures of the pole terms appearing in each GF, this is the minimum number of GFs, which enable us to extract 25 renormalization conditions for the full determination of the mixing matrix. In particular, the GFs that we consider are as follows\footnote{For simplicity of notation, we drop Lorentz and color indices from the GFs; we will reinsert them where needed in the sequel.}:
\begin{enumerate}
\item  Amputated GFs with two external gluon fields and zero-momentum operator insertion, 
\begin{equation}
G_{gi} (q, -q) \equiv \langle A^a_\rho (q) {O_i}_{\mu \nu} (0) A^b_\sigma (-q) \rangle_{\rm amp}, \ (i = 1, \ldots, 5).
\label{Ggi1}
\end{equation} 
\item  Amputated GFs with two external gluon fields and nonzero-momentum operator insertion. For simplicity, we may set to zero the momentum of one of the two external gluons,
\begin{equation}
G_{gi} (q, 0) \equiv \langle A^a_\rho (q) {O_i}_{\mu \nu} (-q) A^b_\sigma (0) \rangle_{\rm amp}, \ (i = 1, \ldots, 5).
\label{Ggi2}
\end{equation}
These GFs are needed to disentangle operator $O_3$ from $O_4$ as they only differ by a total derivative.
\item Amputated GFs with a pair of external quark and antiquark fields and zero-momentum operator insertion,
\begin{equation}
  G_{qi} (q,q) \equiv \langle \psi^{a_f} (q) {O_i}_{\mu \nu} (0) \bar{\psi}^{b_f} (q) \rangle_{\rm amp}, \ (i = 1, \ldots, 5),
  \label{Gqi}
\end{equation}
where $a_f, b_f$ are color indices in the fundamental representation. These GFs are needed to disentangle the fermion operator $O_2$ from the remaining gluon operators. 
\end{enumerate} 
Clearly, the above choices of GFs are not unique; e.g., one can choose to consider GFs with external ghost fields. However, such a choice is not optimal for studying these operators in compact lattice simulations.\footnote{See, however, Refs. \cite{Ghiotti:2007qn, De:2019hov} and references therein for an attempt to address such GFs in lattice simulations.}

\bigskip

As we are interested in calculating GFs with external gluon and quark fields, we also need to compute the renormalization functions of the external fields. To this end, the gluon and quark propagators must also be calculated up to two loops,
\begin{eqnarray}
G_g (q) &\equiv & \langle A^a_\rho (q) A^b_\sigma (-q) \rangle, \\
G_q (q) &\equiv & \langle \psi^{a_f} (q) \bar{\psi}^{b_f} (q) \rangle.
\end{eqnarray} 
Explicit results for these GFs can be found in the literature up to four loops \cite{Ruijl:2017eht}. Also, five-loop results for the renormalization functions of the gluon and quark fields are presented in Ref.~\cite{Chetyrkin:2017bjc}. For completeness, we calculate these GFs up to two loops and we make the crosscheck. A difference between these studies and our work is that we present the conversion factors of the gluon and quark fields between RI$'$ and $\MSbar$ schemes using independent momentum scales. 

\bigskip

\begin{figure}[!htbp] 
  \centering
  \epsfig{file=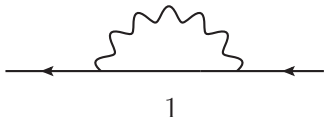,scale=0.8} 
  \caption{One-loop Feynman diagram contributing to the quark propagator $G_q (q)$. The straight (wavy) lines represent fermions (gluons).}
  \label{Fig.FeynmanDiagramsGq}
\end{figure}

\begin{figure}[!htbp] 
  \centering
  \epsfig{file=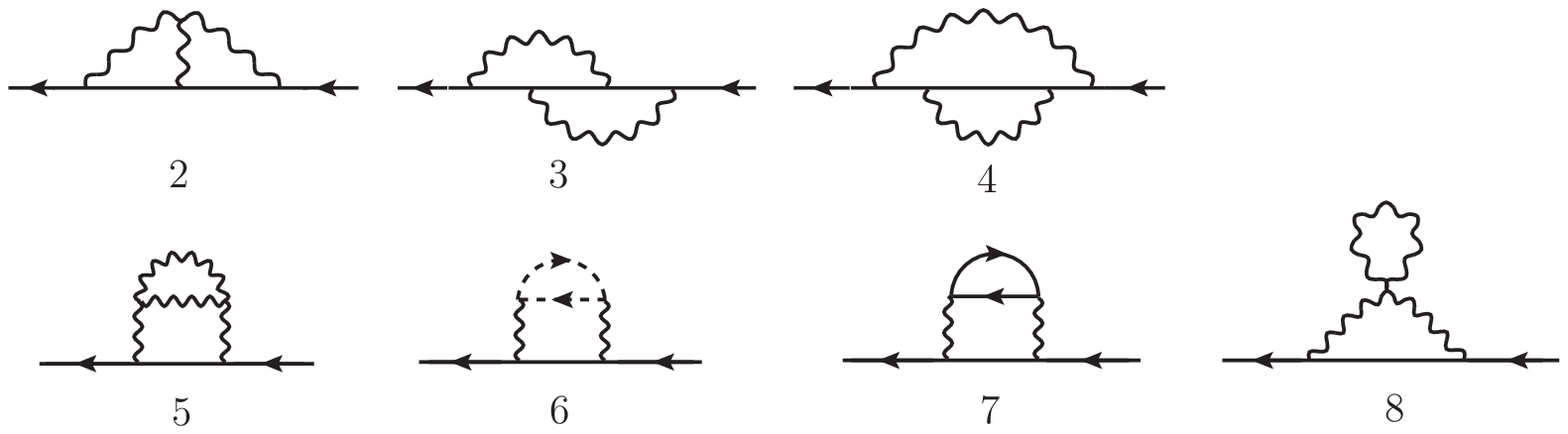} 
  \caption{Two-loop Feynman diagrams contributing to the quark propagator $G_q (q)$. The straight (wavy, dashed) lines represent fermions (gluons, ghosts).} 
 \label{Fig.FeynmanDiagramsGq2}
\end{figure}

\begin{figure}[!htbp] 
  \centering
  \epsfig{file=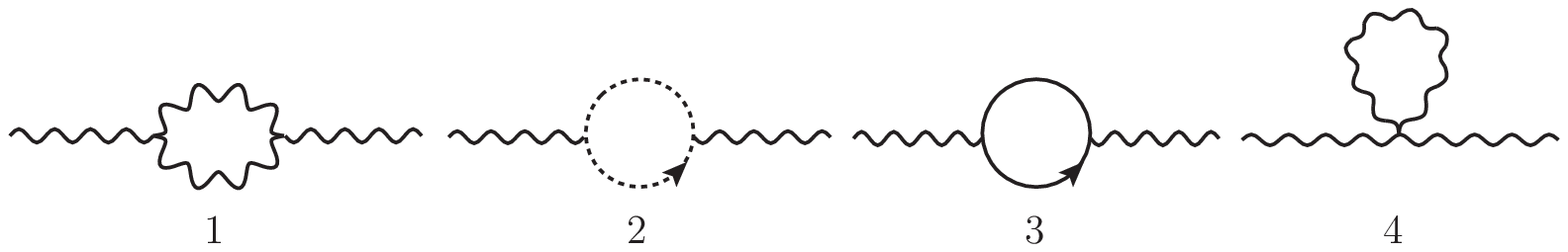} 
  \caption{One-loop Feynman diagrams contributing to the gluon propagator $G_g (q)$. The straight (wavy, dashed) lines represent fermions (gluons, ghosts).} 
 \label{Fig.FeynmanDiagramsGg}
\end{figure}

\begin{figure}[!htbp] 
  \centering
  \epsfig{file=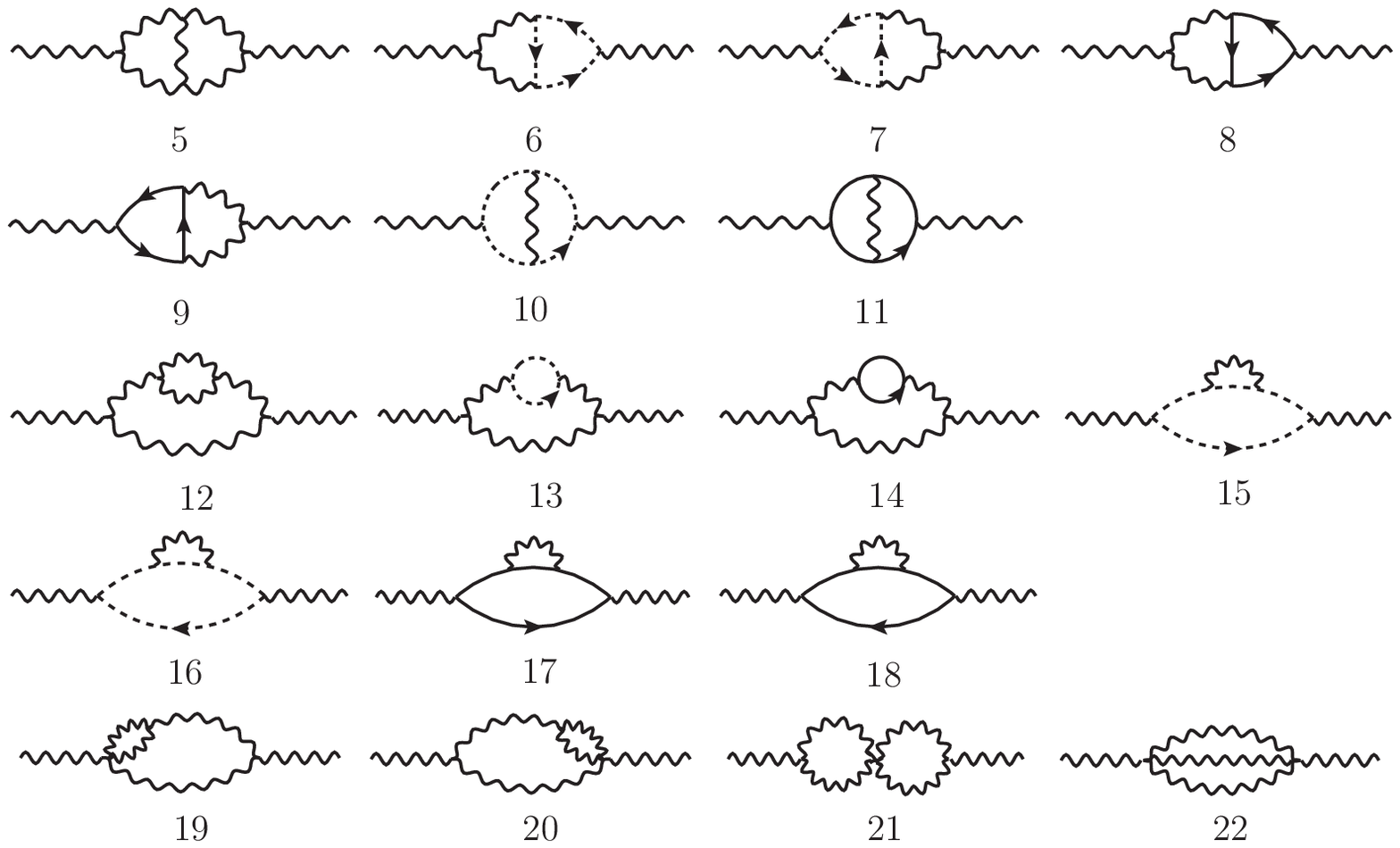} \\
  \hspace{-0.5cm}\epsfig{file=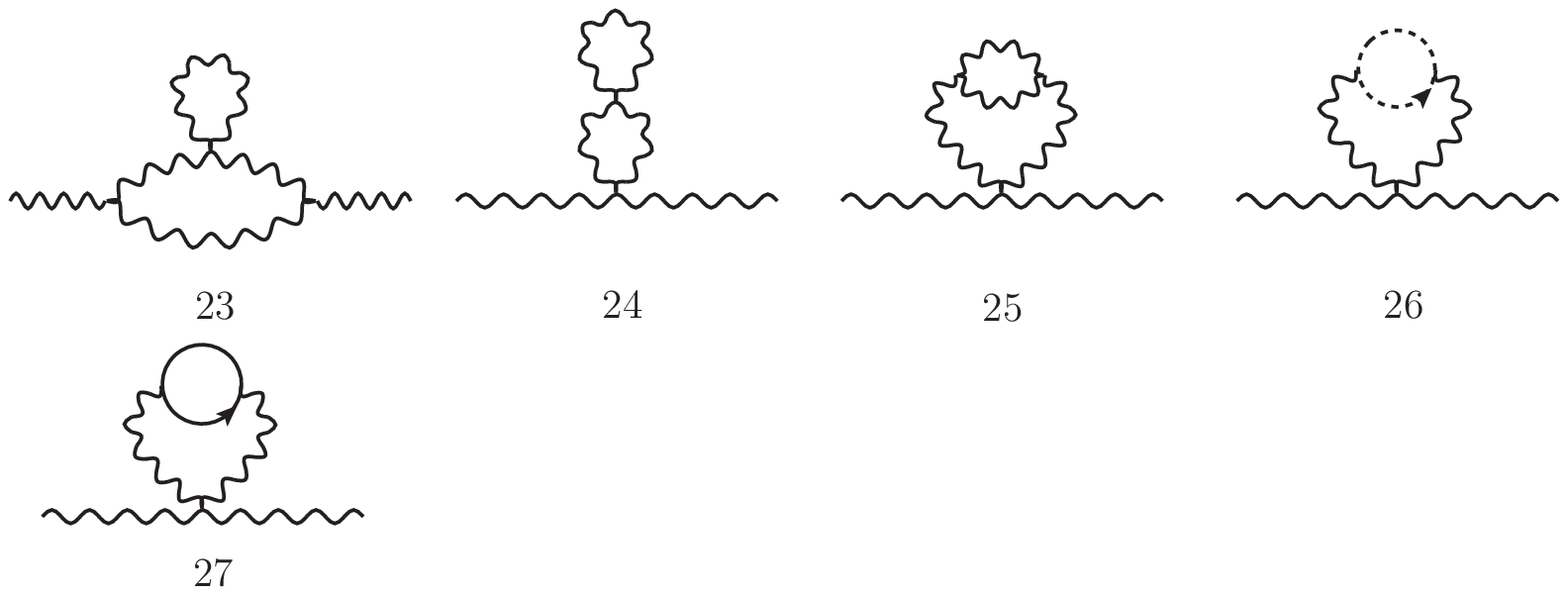,scale=0.97} 
  \caption{Two-loop Feynman diagrams contributing to the gluon propagator $G_g (q)$. The straight (wavy, dashed) lines represent fermions (gluons, ghosts).} 
 \label{Fig.FeynmanDiagramsGg2}
\end{figure}

There are 1 one-loop and 7 two-loop Feynman diagrams contributing to $G_q (q)$, shown in Figs. \ref{Fig.FeynmanDiagramsGq}, \ref{Fig.FeynmanDiagramsGq2}, and 4 one-loop and 23 two-loop Feynman diagrams contributing to $G_g (q)$, shown in Figs. \ref{Fig.FeynmanDiagramsGg} and \ref{Fig.FeynmanDiagramsGg2}. 
The diagrams contributing to $G_{qi} (q, q)$ can be produced by inserting the operator $O_i$ in the vertices or in the propagators of the diagrams of Figs. \ref{Fig.FeynmanDiagramsGq} and \ref{Fig.FeynmanDiagramsGq2}. Similarly, the diagrams of $G_{gi} (q, -q)$ and $G_{gi} (q, 0)$ can be produced from the diagrams of Figs. \ref{Fig.FeynmanDiagramsGg} and \ref{Fig.FeynmanDiagramsGg2} using the same procedure. There is a total of 132, 382, 421 diagrams contributing to $G_{qi} (q,q)$, $G_{gi} (q,-q)$, $G_{gi} (q,0)$, respectively. Note that a number of duplicate diagrams may arise and must not be double-counted.
As is standard practice, we apply the integration by parts method to reduce two-loop integrals into nested one-loop master integrals, which are evaluated by a well-known one-loop formula (see Ref.~\cite{Chetyrkin:1981qh}). The most difficult part of this calculation regards the nonscalar integrands stemming from the ``diamond''-type diagrams (2--3 of Fig. \ref{Fig.FeynmanDiagramsGq2} and 5--11 of Fig. \ref{Fig.FeynmanDiagramsGg2}); we apply an extension of the scalar recursion formula of Ref.~\cite{Chetyrkin:1981qh}, including tensor structures,
\begin{eqnarray}
&\hspace{-1.75cm}(d + n - \alpha_1 - \alpha_2 -2 \alpha_3) \ I_n (\alpha_1, \alpha_2, \alpha_3, \alpha_4, \alpha_5) - \sum_{i=1}^n J_{n-1,i} (\alpha_1, \alpha_2, \alpha_3, \alpha_4, \alpha_5) \nonumber \\
&+ \alpha_1 \Big[ I_n (\alpha_1 + 1, \alpha_2, \alpha_3, \alpha_4-1, \alpha_5) - I_n (\alpha_1 + 1, \alpha_2, \alpha_3-1, \alpha_4, \alpha_5) \Big] \nonumber \\
&\qquad \ \ + \alpha_2 \Big[ I_n (\alpha_1, \alpha_2 + 1, \alpha_3, \alpha_4, \alpha_5-1) - I_n (\alpha_1, \alpha_2 + 1, \alpha_3-1, \alpha_4, \alpha_5) \Big]  = 0, \nonumber \\
& \qquad \qquad \qquad \qquad \qquad \qquad \qquad \qquad \quad (\alpha_3 \in \mathbb{Z}^+, \alpha_4 \in \mathbb{Z}^+, \alpha_5 \in \mathbb{Z}^+, n \in \mathbb{Z}^+),
\label{diamond_int} 
\end{eqnarray} 
where
\begin{eqnarray}
I_n (\alpha_1, \alpha_2, \alpha_3, \alpha_4, \alpha_5) &\equiv & \int \frac{d^d p}{{(2 \pi)}^d} \int \frac{d^d k}{{(2 \pi)}^d} \frac{f(k) \ p_{\mu_1} \ldots p_{\mu_n}}{p^{2 \alpha_1} {(p - q)}^{2 \alpha_2} {(p-k)}^{2 \alpha_3} k^{2 \alpha_4} {(k - q)}^{2 \alpha_5}}, \ \ \\
J_{n-1,i} (\alpha_1, \alpha_2, \alpha_3, \alpha_4, \alpha_5) &\equiv & \int \frac{d^d p}{{(2 \pi)}^d} \int \frac{d^d k}{{(2 \pi)}^d} \frac{f(k) \ p_{\mu_1} \ldots p_{\mu_n} \cdot (k_{\mu_i} / p_{\mu_i})}{p^{2 \alpha_1} {(p - q)}^{2 \alpha_2} {(p-k)}^{2 \alpha_3} k^{2 \alpha_4} {(k - q)}^{2 \alpha_5}}, \ \
\end{eqnarray}
$f (k)$ is a function of $k$, and $q$ is an external momentum $d$-vector. For $n=0$, Eq. \eqref{diamond_int} reduces to the following scalar formula:
\begin{eqnarray}
&\hspace{-8cm}(d - \alpha_1 - \alpha_2 -2 \alpha_3) \ I_0 (\alpha_1, \alpha_2, \alpha_3, \alpha_4, \alpha_5)  \nonumber \\
&+ \alpha_1 \Big[ I_0 (\alpha_1 + 1, \alpha_2, \alpha_3, \alpha_4-1, \alpha_5) - I_0 (\alpha_1 + 1, \alpha_2, \alpha_3-1, \alpha_4, \alpha_5) \Big] \nonumber \\
&\qquad \ \ + \alpha_2 \Big[ I_0 (\alpha_1, \alpha_2 + 1, \alpha_3, \alpha_4, \alpha_5-1) - I_0 (\alpha_1, \alpha_2 + 1, \alpha_3-1, \alpha_4, \alpha_5) \Big]  = 0, \nonumber \\
& \qquad \qquad \qquad \qquad \qquad \qquad \qquad \qquad \qquad \qquad \ \ (\alpha_3 \in \mathbb{Z}^+, \alpha_4 \in \mathbb{Z}^+, \alpha_5 \in \mathbb{Z}^+),
\label{diamond_int2} 
\end{eqnarray} 
where
\begin{equation}
I_0 (\alpha_1, \alpha_2, \alpha_3, \alpha_4, \alpha_5) \equiv \int \frac{d^d p}{{(2 \pi)}^d} \int \frac{d^d k}{{(2 \pi)}^d} \frac{f(k)}{p^{2 \alpha_1} {(p - q)}^{2 \alpha_2} {(p-k)}^{2 \alpha_3} k^{2 \alpha_4} {(k - q)}^{2 \alpha_5}}.
\end{equation}
Another possibility is to express all integrals in terms of scalar functions of the external momentum by multiplying each integral with the appropriate projectors.
\subsection{Renormalization schemes and conversion factors}

In our study, we adopt two different renormalization schemes: the $\MSbar$ scheme, which is typically used in phenomenology for the analysis of experimental data, and a RI$'$ scheme, which is more immediate for a lattice regularized theory. The latter scheme is appropriate for renormalizing nonperturbative data taken by lattice simulations. Given that $\MSbar$ is defined in a perturbative manner, the best theoretical approach for taking nonperturbative results in $\MSbar$ is to make use of an intermediate scheme, which is applicable in both perturbative and nonperturbative regularizations, and to match the nonperturbative results from this scheme to $\MSbar$; RI$'$ is an example of such an intermediate scheme. RI$'$-renormalized quantities, calculated on the lattice nonperturbatively, can be converted to the $\MSbar$ counterparts through perturbative ``conversion'' factors between RI$'$ and $\MSbar$ schemes; the conversion factors are regularization independent and thus, calculable in DR.

\bigskip

Below, we provide our conventions for the definition of the renormalization functions, which relate bare to renormalized fields and parameters of the theory,
\begin{eqnarray}
A_\mu^Y &\equiv & {\left(Z_A^{Y,X}\right)}^{-1/2} A_\mu^X, \label{AR}\\
\psi_f^Y &\equiv & {\left(Z_\psi^{Y,X}\right)}^{-1/2} \psi_f^X, \\
g^Y &\equiv & \mu^{(D - 4)/2} {\left(Z_g^{Y,X}\right)}^{-1} g^X, \\
\alpha^Y &\equiv & Z_\alpha^{Y,X} {\left(Z_A^{Y,X}\right)}^{-1} \alpha^X, \label{alphaR}
\end{eqnarray}
where $A_\mu$ is the gluon field, $\psi_f$ is the quark field of flavor $f$, $g$ is the coupling constant, $\alpha$ is the gauge-fixing parameter ($\alpha = 0$ in the Landau gauge), and $\mu$ is a momentum scale. The index $X$ denotes bare quantities in the $X$ regularization, and the index $Y$ denotes renormalized quantities in the $Y$ renormalization scheme. The $\MSbar$ renormalization scale $\bar{\mu}$ is defined in terms of $\mu$,
\begin{equation}
\bar{\mu} \equiv \mu {\left(\frac{4 \pi}{e^{\gamma_E}}\right)}^{1/2},
\end{equation} 
where $\gamma_E$ is Euler's gamma. The renormalization functions for the operators under study have been already defined in Eq. \eqref{Z_O} in a $5 \times 5$ matrix form. The renormalized Green's functions of operators and fields, which are defined in the previous subsection, are given by
\begin{eqnarray}
G_{g}^Y &=& {\left(Z^{Y,X}_A\right)}^{-1} G_{g}^X, \\
G_{q}^Y &=& {\left(Z^{Y,X}_\psi\right)}^{-1} G_{q}^X, \\
G_{gi}^Y &=& Z^{Y,X}_A \sum_{j = 1}^5 Z^{Y,X}_{ij} G_{gi}^X, \quad (i=1, \ldots, 5), \\
G_{qi}^Y &=& Z^{Y,X}_\psi \sum_{j = 1}^5 Z^{Y,X}_{ij} G_{qi}^X, \quad (i=1, \ldots, 5).
\end{eqnarray}

\bigskip

In the $\MSbar$ scheme, the renormalization condition is defined (in DR) by imposing that the renormalized Green's functions are finite, when the renormalization functions include only negative powers of $\varepsilon \equiv (4 - d)/2$. In a RI$'$-like scheme, there is, {\it a priori}, wide flexibility in defining normalization conditions in Green's functions, especially when operator mixing is present. The possible variants differ only by finite terms. Therefore, it is natural to adopt a minimal prescription, which involves the smallest possible set of operators which can mix; this is usually the mixing set found in $\MSbar$. Of course, the conditions must be regularization independent and thus, they must also include any possible additional finite or power-divergent mixing, which is present, e.g., in the lattice regularization. Examples of operators with additional mixing on the lattice are the scalar glueball operator, the scalar quark-antiquark operator, as well as the nonlocal quark bilinears \cite{Constantinou:2017sej,Spanoudes:2018zya,Constantinou:2019vyb} studied in a chiral-symmetry breaking action. In the present case, such admixtures on the lattice are excluded by hypercubic invariance.     

\bigskip

A choice of definition for a RI$'$-like scheme, compatibly with $\MSbar$, is to consider a $5 \times 5$ mixing matrix. The elements of the mixing matrix are obtained by imposing $5 \times 5 = 25$ conditions on Green's functions. This can be done by isolating different Lorentz and Dirac structures of each Green's function. Given that the operators under study are two-index $(\mu, \nu)$ symmetric, the possible structures for the Green's functions under study with external gluon fields $A_\rho (q)$, $A_\sigma (-q)$ [or $A_\rho (q), A_\sigma (0)$, cf. Eqs. (\ref{Ggi1}, \ref{Ggi2})] are (for $\mu \neq \nu$),
\begin{eqnarray}
&& \delta_{\rho \sigma} q_\mu q_\nu, \qquad q^2 (\delta_{\rho \mu} \delta_{\sigma \nu} + \delta_{\rho \nu} \delta_{\sigma \mu}), \qquad (q_\rho q_\mu \delta_{\sigma \nu} + q_\rho q_\nu \delta_{\sigma \mu}), \\
&& \qquad \qquad (q_\sigma q_\mu \delta_{\rho \nu} + q_\sigma q_\nu \delta_{\rho \mu}), \qquad q_\mu q_\nu q_\rho q_\sigma / q^2. 
\end{eqnarray}
Similarly, the possible structures for the fermionic Green's functions under study are (for $\mu \neq \nu$) as follows:
\begin{equation}
(\gamma_\mu q_\nu + \gamma_\nu q_\mu), \qquad \slashed{q} q_\mu q_\nu / q^2.
\end{equation}
We isolate some of these structures, including those with poles, by selecting specific values for the external momentum and/or the Lorentz components of the external fields; for these specific values, we impose that
\begin{equation}
{\rm Tr} \left[ G_{gi} \right] = {\rm Tr} \left[ G_{gi}^{\rm tree} \right], 
\end{equation}
and similarly
\begin{equation}
{\rm Tr} \left[ G_{qi} \cdot \slashed{q} \right] = {\rm Tr} \left[ G_{qi}^{\rm tree} \cdot \slashed{q} \right].
\end{equation}
The proposed renormalization conditions for this variant of RI$'$, dubbed ${\rm RI}_1'$, are [cf. Eqs. (\ref{Ggi1} -- \ref{Gqi})]
\begin{equation}
\frac{{\rm Tr} [G_{gi}^{{\rm RI}_1'} (q,-q)]}{N_c^2 - 1} \Bigg\vert_{\begin{smallmatrix}
\rho = \sigma, \\
\rho \neq (\mu, \nu), \\
q_\rho = 0, \\
q_\tau = \bar{q}_\tau, \ \forall \tau \neq \rho
\end{smallmatrix}} = \frac{{\rm Tr} [G_{gi}^{\rm tree} (q,-q)]}{N_c^2 - 1} \Bigg\vert_{\begin{smallmatrix}
\rho = \sigma, \\
\rho \neq (\mu, \nu), \\
q_\rho = 0, \\
q_\tau = \bar{q}_\tau, \ \forall \tau \neq \rho
\end{smallmatrix}} = \Bigg\{ \begin{matrix}
\ 2 \bar{q}_\mu \bar{q}_\nu, \ &i=1 \\
\\
0, \ &i = 2, 3, 4, 5,
\end{matrix} \label{RI1_1}
\end{equation}
\begin{equation}
\frac{{\rm Tr} [G_{gi}^{{\rm RI}_1'} (q,-q)]}{N_c^2 - 1} \Bigg\vert_{\begin{smallmatrix}
\rho = \mu, \\
\sigma = \nu, \\
q_\rho = q_\sigma = 0, \\
q_\tau = \bar{q}_\tau, \ \forall \tau \neq (\rho, \sigma)
\end{smallmatrix}} = \frac{{\rm Tr} [G_{gi}^{\rm tree} (q,-q)]}{N_c^2 - 1} \Bigg\vert_{\begin{smallmatrix}
\rho = \mu, \\
\sigma = \nu, \\
q_\rho = q_\sigma = 0, \\
q_\tau = \bar{q}_\tau, \ \forall \tau \neq (\rho, \sigma)
\end{smallmatrix}} = \Bigg\{ \begin{matrix}
\ \bar{q}^2, \ &i=1 \\
0, \ &i = 2, 3, 4 \\
\ 2 \bar{q}^2, \ &i=5,
\end{matrix} \label{RI1_2}
\end{equation}
\begin{equation}
\frac{{\rm Tr} [G_{gi}^{{\rm RI}_1'} (q,-q)]}{N_c^2 - 1} \Bigg\vert_{\begin{smallmatrix}
\rho \neq (\mu, \nu), \\
\sigma = \nu, \\
q_\sigma = 0, \\
q_\tau = \bar{q}_\tau, \ \forall \tau \neq \sigma
\end{smallmatrix}} = \frac{{\rm Tr} [G_{gi}^{\rm tree} (q,-q)]}{N_c^2 - 1} \Bigg\vert_{\begin{smallmatrix}
\rho \neq (\mu, \nu), \\
\sigma = \nu, \\
q_\sigma = 0, \\
q_\tau = \bar{q}_\tau, \ \forall \tau \neq \sigma
\end{smallmatrix}} = \Bigg\{ \begin{matrix}
\ - \bar{q}_\mu \bar{q}_\rho, \ &i=1 \\
0, \ &i = 2 \\
\ \bar{q}_\mu \bar{q}_\rho / \alpha^{{\rm RI}'}, \ &i=3, 4 \\
\ \bar{q}_\mu \bar{q}_\rho \left(1 / \alpha^{{\rm RI}'} - 1 \right), \ &i=5,
\end{matrix} \label{RI1_3}
\end{equation}
\begin{equation}
\frac{{\rm Tr} [G_{gi}^{{\rm RI}_1'} (q,0)]}{N_c^2 - 1}  \Bigg\vert_{\begin{smallmatrix}
\rho \neq (\mu, \nu), \\
\sigma = \nu, \\
q_\sigma = 0, \\
q_\tau = \bar{q}_\tau, \ \forall \tau \neq \sigma
\end{smallmatrix}} = \frac{{\rm Tr} [G_{gi}^{\rm tree} (q,0)]}{N_c^2 - 1} \Bigg\vert_{\begin{smallmatrix}
\rho \neq (\mu, \nu), \\
\sigma = \nu, \\
q_\sigma = 0, \\
q_\tau = \bar{q}_\tau, \ \forall \tau \neq \sigma
\end{smallmatrix}} = \Bigg\{ \begin{matrix}
0, \ &i = 1, 2, 3 \\
\ \bar{q}_\mu \bar{q}_\rho / \alpha^{{\rm RI}'}, \ &i=4 \\
\ \bar{q}_\mu \bar{q}_\rho \left(1 / \alpha^{{\rm RI}'} - 1 \right), \ &i=5,
\end{matrix} \label{RI1_4}
\end{equation}
\begin{equation}
\frac{1}{4 N_c} {\rm Tr} [G_{qi}^{{\rm RI}_1'} (q, q) \cdot \slashed{q}] \Bigg\vert_{\begin{smallmatrix}
q_\tau = \bar{q}_\tau, \forall \tau
\end{smallmatrix}} = \frac{1}{4 N_c} {\rm Tr} [G_{qi}^{\rm tree} (q,q) \cdot \slashed{q}] \Bigg\vert_{\begin{smallmatrix}
q_\tau = \bar{q}_\tau, \forall \tau
\end{smallmatrix}} = \Bigg\{ \begin{matrix}
0, \ &i = 1, 3, 4, 5 \\
\\
\ i \bar{q}_\mu \bar{q}_\nu, \ &i=2,
\end{matrix} \label{RI1_5}
\end{equation}
where the trace in Eqs. (\ref{RI1_1} -- \ref{RI1_4}) is taken over color space (in the adjoint representation), and the trace in Eq. \eqref{RI1_5} is taken over Dirac and color spaces (in the fundamental representation); the four-vector $\bar{q}$ is the RI$'$ renormalization scale.

\bigskip

The above prescription is not a minimal one. From our two-loop results, one can observe that the mixing pattern in the $\MSbar$ scheme reduces to a set of three operators: $\{ O_1, O_2, O_6 \equiv O_4 - O_5 \}$. This was expected from the theoretical analysis presented in Sec. \ref{TheoreticalAnalysis}. Thus, a second choice of definition for a RI$'$-like scheme is to consider a $3 \times 3$ mixing matrix. Now, we only need nine conditions to identify the renormalization factors. The first two and the last condition of the ${\rm RI}_1'$ scheme (Eqs. \ref{RI1_1}, \ref{RI1_2}, \ref{RI1_5}) taken for the three operators $\{ O_1, O_2, O_6 \}$ can be also the conditions for the ${\rm RI}_2'$ scheme,
\begin{equation}
\frac{{\rm Tr} [G_{gi}^{{\rm RI}_2'} (q,-q)]}{N_c^2 - 1} \Bigg\vert_{\begin{smallmatrix}
\rho = \sigma, \\
\rho \neq (\mu, \nu), \\
q_\rho = 0, \\
q_\tau = \bar{q}_\tau, \ \forall \tau \neq \rho
\end{smallmatrix}} = \frac{{\rm Tr} [G_{gi}^{\rm tree} (q,-q)]}{N_c^2 - 1} \Bigg\vert_{\begin{smallmatrix}
\rho = \sigma, \\
\rho \neq (\mu, \nu), \\
q_\rho = 0, \\
q_\tau = \bar{q}_\tau, \ \forall \tau \neq \rho
\end{smallmatrix}} = \Bigg\{ \begin{matrix}
\ 2 \bar{q}_\mu \bar{q}_\nu, \ &i=1 \\
\\
0, \ &i = 2, 6,
\end{matrix} \label{RI2_1}
\end{equation}
\begin{equation}
\frac{{\rm Tr} [G_{gi}^{{\rm RI}_2'} (q,-q)]}{N_c^2 - 1} \Bigg\vert_{\begin{smallmatrix}
\rho = \mu, \\
\sigma = \nu, \\
q_\rho = q_\sigma = 0, \\
q_\tau = \bar{q}_\tau, \ \forall \tau \neq (\rho, \sigma)
\end{smallmatrix}} = \frac{{\rm Tr} [G_{gi}^{\rm tree} (q,-q)]}{N_c^2 - 1} \Bigg\vert_{\begin{smallmatrix}
\rho = \mu, \\
\sigma = \nu, \\
q_\rho = q_\sigma = 0, \\
q_\tau = \bar{q}_\tau, \ \forall \tau \neq (\rho, \sigma)
\end{smallmatrix}} = \Bigg\{ \begin{matrix}
\ \bar{q}^2, \ &i=1 \\
0, \ &i = 2 \\
\ - 2 \bar{q}^2, \ &i=6,
\end{matrix} \label{RI2_2}
\end{equation}
\begin{equation}
\frac{1}{4 N_c} {\rm Tr} [G_{qi}^{{\rm RI}_2'} (q, q) \cdot \slashed{q}] \Bigg\vert_{\begin{smallmatrix}
q_\tau = \bar{q}_\tau, \forall \tau
\end{smallmatrix}} = \frac{1}{4 N_c} {\rm Tr} [G_{qi}^{\rm tree} (q,q) \cdot \slashed{q}] \Bigg\vert_{\begin{smallmatrix}
q_\tau = \bar{q}_\tau, \forall \tau
\end{smallmatrix}} = \Bigg\{ \begin{matrix}
0, \ &i = 1, 6 \\
\\
\ i \bar{q}_\mu \bar{q}_\nu, \ &i=2.
\end{matrix} \label{RI2_3}
\end{equation}
This scheme has the advantage of not involving GFs with nonzero momentum operator insertions.

\bigskip

A third choice for defining a RI$'$-like scheme is to impose that the EMT, which is constructed by $O_1$, $O_2$, $O_4$, and/or $O_5$, is still a conserved quantity after its renormalization in RI$'$ scheme. In DR, the conservation gives $O_1^{{\rm RI}'} + O_2^{{\rm RI}'} + O_4^{{\rm RI}'} = O_1^{\rm DR} + O_2^{\rm DR} + O_4^{\rm DR}$ and Eqs. (\ref{cons1} -- \ref{cons5}) will also hold to this version of RI$'$. As we insert five new conditions, we must exclude five conditions from the previous definition of ${\rm RI}'_1$ scheme. For example, we exclude the operator $O_4$ from each condition [Eqs. (\ref{RI1_1} -- \ref{RI1_5})]. Similarly, we can define the ``conserved'' version of the ${\rm RI}'_2$ scheme. On the lattice, the construction of a conserved EMT is more complex due to the presence of discretization effects, which violate translational invariance. A discussion about the possible ways of applying the conservation properties of EMT on the lattice is given in Sec. \ref{Nonperturbative renormalization}.

\bigskip

The above variants of RI$'$ involve operator Green's functions of exceptional momentum configurations in the sense of having one out of three external momenta set to zero and, thus, it is important to make sure that no infrared (IR) divergences arise as a consequence; at the perturbative level, we have verified in all our two-loop calculations that no such divergence is present in any diagram. Nevertheless, such Green's functions have IR issues on the lattice in the chiral (massless) limit. Thus, it is a challenge to choose a range of low momenta in the nonperturbative studies, which leads to reliable results in that limit. IR issues attached to the chiral limit are under better control when studying Green's functions with nonexceptional momentum configurations (see, e.g.,~\cite{Sturm:2009kb}). A potential variant of RI$'$, including nonexceptional momenta, is the RI/SMOM scheme (regularization-independent symmetric momentum-subtraction scheme; for an application, see, e.g.,~\cite{Bi:2017ybi}). The computation of conversion factors between this scheme and $\MSbar$ proceeds in an analogous fashion to what is presented in Sec.~\ref{Conversion factors}, but it is beyond the scope of this work.

\bigskip

To complete the renormalization prescription, we also provide the conditions for the RI$'$ renormalization factors of gluon and fermion fields,
\begin{eqnarray}
\frac{1}{N_c^2 - 1} \frac{1}{d - 1} \sum_{\rho, \sigma} {\rm Tr} \left[{\left( G_g^{{\rm RI}'} (q) \right)}_{\rho \sigma} \cdot \left(q^2 \delta_{\rho \sigma} - q_\rho q_\sigma \right)\right] \Big\vert_{q^2 = \bar{q}^2} &=& 1, \label{RI_A}\\
\frac{1}{4 N_c} {\rm Tr} {\left[G_q^{{\rm RI}'} (q) \cdot (i \slashed{q})\right]} \Big\vert_{q^2 = \bar{q}^2} &=& 1, \label{RI_psi}
\end{eqnarray}
where the trace in Eq. \eqref{RI_A} is taken over color space (in the adjoint representation), and the trace in Eq. \eqref{RI_psi} is taken over Dirac and color spaces (in the fundamental representation).

\bigskip

Finally, the passage to the $\MSbar$ scheme can be achieved by using the conversion factors between the different versions of RI$'$ and the $\MSbar$ scheme, defined as
\begin{eqnarray}
C_{ij}^{\MSbar, {\rm RI}'} & \equiv & \sum_k Z_{ik}^{\MSbar, {\rm LR}} {\left[{\left(Z^{{\rm RI}', {\rm LR}}\right)}^{-1}\right]}_{kj} = \sum_k Z_{ik}^{\MSbar, {\rm DR}} {\left[{\left(Z^{{\rm RI}', {\rm DR}}\right)}^{-1} \right]}_{kj}, \label{conv1} \\
C_A^{\MSbar, {\rm RI}'} & \equiv & Z_A^{\MSbar, {\rm LR}} / Z_A^{{\rm RI}', {\rm LR}} = Z_A^{\MSbar, {\rm DR}} / Z_A^{{\rm RI}', {\rm DR}}, \\ 
C_\psi^{\MSbar, {\rm RI}'} & \equiv & Z_\psi^{\MSbar, {\rm LR}} / Z_\psi^{{\rm RI}', {\rm LR}} = Z_\psi^{\MSbar, {\rm DR}} / Z_\psi^{{\rm RI}', {\rm DR}},
\end{eqnarray} 
for the set of mixing operators, the gluon and the quark field, respectively. Note that in Eq. \eqref{conv1}, $i,j,k = 1, 2, 3, 4, 5$ for ${\rm RI_1}'$, and $i,j,k = 1, 2, 6$ for ${\rm RI_2}'$; also, the superscript ${\rm LR}$ means Lattice regularization. 

\section{Results}
\label{Results}

In this section, we present our one- and two-loop results for the $\MSbar$-renormalized Green's functions of the operators under study, the renormalization factors and the conversion factors between the different RI$'$ versions and the $\MSbar$ scheme, which are all described in the previous section. To facilitate the use of all these results, we provide them also as Supplemental Material \cite{SupMat}, in the form of two equivalent files: a {\it Mathematica} input file: ``Greens\_Functions\_and\_Conversion\_Factors.m'', and a text version of the same file: ``Greens\_Functions\_and\_Conversion\_Factors.txt''.

\subsection{$\MSbar$-renormalized Green's functions}

Here, we provide our resulting expressions for the $\MSbar$-renormalized Green's functions of operators $O_i$, $(i=1, 2, \ldots, 5)$ in terms of the following combinations of Lorentz and Dirac structures\footnote{For notation, see Sec. \ref{Calculation Setup}.}:
\begin{eqnarray}
G_1 &\equiv & G^{\rm tree}_{g1} (q,-q) = \delta^{ab} \Big( 2 q_\mu q_\nu \delta_{\rho \sigma} + q^2 \left( \delta_{\rho \mu} \delta_{\sigma \nu} + \delta_{\rho \nu} \delta_{\sigma \mu} \right) - \left(q_\rho q_\mu \delta_{\sigma \nu} + q_\rho q_\nu \delta_{\sigma \mu} \right) \nonumber \\
&& \qquad \qquad \qquad \qquad \ - \left(q_\sigma q_\mu \delta_{\rho \nu} + q_\sigma q_\nu \delta_{\rho \mu} \right) \Big), \\
G_2 &\equiv & G^{\rm tree}_{g3} (q,-q) = G^{\rm tree}_{g4} (q,-q) = \delta^{ab} \frac{1}{\alpha_\MSbar} \Big( \left(q_\rho q_\mu \delta_{\sigma \nu} + q_\rho q_\nu \delta_{\sigma \mu} \right) \nonumber \\
&& \qquad \qquad \qquad \qquad \qquad \qquad \qquad \qquad \quad + \left(q_\sigma q_\mu \delta_{\rho \nu} + q_\sigma q_\nu \delta_{\rho \mu} \right)\Big), \\
G_3 &\equiv & G^{\rm tree}_{g5} (q,-q) =  \delta^{ab} \Big( 2 q^2 \left( \delta_{\rho \mu} \delta_{\sigma \nu} + \delta_{\rho \nu} \delta_{\sigma \mu} \right) + \nonumber \\ 
&& \qquad \qquad \qquad \left( \frac{1}{\alpha_\MSbar} - 1 \right) \Big( \left(q_\rho q_\mu \delta_{\sigma \nu} + q_\rho q_\nu \delta_{\sigma \mu} \right) + \left(q_\sigma q_\mu \delta_{\rho \nu} + q_\sigma q_\nu \delta_{\rho \mu} \right)\Big)\Big), \\
G_4 &\equiv & \delta^{ab} q_\mu q_\nu q_\rho q_\sigma / q^2, \\
G_5 &\equiv & G^{\rm tree}_{g4} (q,0) = \delta^{ab} \frac{1}{\alpha_\MSbar} \Big( q_\rho q_\mu \delta_{\sigma \nu} + q_\rho q_\nu \delta_{\sigma \mu} \Big), \\
G_6 &\equiv & G^{\rm tree}_{g5} (q,0) =  \delta^{ab} \Big( q^2 \left( \delta_{\rho \mu} \delta_{\sigma \nu} + \delta_{\rho \nu} \delta_{\sigma \mu} \right) + \left( \frac{1}{\alpha_\MSbar} - 1 \right) \Big( q_\rho q_\mu \delta_{\sigma \nu} + q_\rho q_\nu \delta_{\sigma \mu} \Big)\Big), \\
G_7 &\equiv & \delta^{ab} \frac{1}{\alpha_\MSbar} \Big( q_\sigma q_\mu \delta_{\rho \nu} + q_\sigma q_\nu \delta_{\rho \mu} \Big), \\
G_8 &\equiv & G^{\rm tree}_{q2} (q,q) = \delta^{a_f b_f} \frac{i}{2} \Big( \gamma_\mu q_\nu + \gamma_\nu q_\mu \Big), \\
G_9 &\equiv & \delta^{a_f b_f} i \frac{\slashed{q} q_\mu q_\nu}{q^2}.
\end{eqnarray}
In what follows, $C_F \equiv (N_c^2 - 1) / (2 N_c)$ is the Casimir operator in the fundamental representation and $\zeta (n)$ is the Riemann zeta function. 

\bigskip

The expressions for the Green's functions with two external gluon fields and zero-momentum operator insertion ($G^\MSbar_{gi} (q,-q)$) are as follows: \\
\begin{eqnarray}
&& G^\MSbar_{g1} (q,-q) = \qquad \qquad \qquad \qquad \qquad \qquad \qquad \qquad \qquad \qquad \qquad \qquad\qquad \qquad \qquad \qquad \quad \nonumber \\
&& \qquad G_1 \Bigg\{ 1 + \frac{g_\MSbar^2}{16 \pi^2} N_c \ \Bigg[- \frac{41}{18} - 2 \alpha_\MSbar - \frac{1}{2} \alpha_\MSbar^2 + \left(\frac{13}{6} - \frac{1}{2} \alpha_\MSbar \right) \logq \Bigg] \nonumber \\
&& \qquad \ \ + \frac{g_\MSbar^4}{(16 \pi^2)^2} \Bigg[ N_c^2 \Bigg(-\frac{46987}{1440} - \frac{2347}{1440} \alpha_\MSbar + \frac{1703}{720} \alpha_\MSbar^2 +\frac{5}{8} \alpha_\MSbar^3 + \frac{3}{16} \alpha_\MSbar^4  + \nonumber \\
&& \qquad \qquad \qquad \qquad \qquad \ \ \ \left(\frac{119}{12} - \frac{149}{72} \alpha_\MSbar + \frac{5}{24} \alpha_\MSbar^2 + \frac{3}{4} \alpha_\MSbar^3 \right) \logq + \nonumber \\
&& \qquad \qquad \qquad \qquad \qquad \ \ \ \left(- \frac{13}{8} - \frac{17}{24} \alpha_\MSbar + \frac{1}{4} \alpha_\MSbar^2 \right) \logqsq + \nonumber \\
&& \qquad \qquad \qquad \qquad \qquad \ \ \ \left(\frac{99}{10} - \frac{3}{5} \alpha_\MSbar - \frac{1}{5} \alpha_\MSbar^2 \right) \zeta (3) \Bigg) + \nonumber \\
&& \qquad \qquad \qquad \qquad \ \ \frac{N_f}{N_c} \Bigg(-\frac{311}{324} + \frac{2}{9} \logq - \frac{4}{9} \logqsq \Bigg) + \nonumber \\
&& \qquad \qquad \qquad \quad \ \ N_f N_c \Bigg(\frac{73}{8} - \frac{5}{9} \alpha_\MSbar^2 + \left( -\frac{53}{18} + \frac{1}{3} \alpha_\MSbar^2 \right) \logq + \nonumber \\
&& \qquad \qquad \qquad \qquad \qquad \ \ \ \ \frac{7}{6} \logqsq - 6 \zeta (3) \Bigg) \Bigg] + \mathcal{O} (g_\MSbar^6) \Bigg\} + \nonumber \\
&& \qquad G_2 \Bigg\{\frac{g_\MSbar^2}{16 \pi^2} N_c \ \left(\frac{5}{6} - \frac{1}{6} \alpha_\MSbar - \frac{1}{2} \alpha_\MSbar^2 - \frac{1}{2} \logq \right) \qquad \qquad \qquad \qquad \qquad \qquad \qquad \qquad \qquad \qquad \nonumber \\
&& \qquad \qquad + \frac{g_\MSbar^4}{(16 \pi^2)^2} \Bigg[ N_c^2 \Bigg(\frac{29}{30} - \frac{1733}{720} \alpha_\MSbar - \frac{859}{1440} \alpha_\MSbar^2 +\frac{3}{5} \alpha_\MSbar^3 + \frac{1}{4} \alpha_\MSbar^4  + \nonumber \\
&& \qquad \qquad \qquad \qquad \quad \ \ \ \ \ \left(-\frac{23}{18} - \frac{41}{24} \alpha_\MSbar + \frac{1}{2} \alpha_\MSbar^3 \right) \logq + \nonumber \\
&& \qquad \qquad \qquad \qquad \qquad \ \ \left(- \frac{7}{24} + \frac{1}{4} \alpha_\MSbar \right) \logqsq + \nonumber \\
&& \qquad \qquad \qquad \qquad \quad \ \ \ \ \ \left(\frac{47}{20} + \frac{123}{40} \alpha_\MSbar + \frac{17}{40} \alpha_\MSbar^2 - \frac{1}{20} \alpha_\MSbar^3 \right) \zeta (3) \Bigg) + \nonumber \\
&& \qquad \qquad \qquad \quad \ \ \ \ \ \frac{N_f}{N_c} \Bigg(-\frac{2}{9} - \frac{2}{9} \alpha_\MSbar + \left( -\frac{4}{9} - \frac{4}{9} \alpha_\MSbar \right) \logq \Bigg) + \nonumber \\
&& \qquad \qquad \qquad \ \ \ \ \ N_f N_c \Bigg(-\frac{157}{108} + \frac{11}{18} \alpha_\MSbar + \left( \frac{11}{9} + \frac{2}{3} \alpha_\MSbar \right) \logq + \nonumber \\
&& \qquad \qquad \qquad \qquad \qquad \ \ \ \ \left(- 2 - 2 \alpha_\MSbar \right) \zeta (3) \Bigg) \Bigg] + \mathcal{O} (g_\MSbar^6) \Bigg\} + \nonumber
\end{eqnarray}
\begin{eqnarray}
&& \qquad G_3 \Bigg\{\frac{g_\MSbar^2}{16 \pi^2} N_c \ \left(- \frac{5}{6} + \frac{1}{2} \alpha_\MSbar + \frac{1}{2} \logq \right) \nonumber \\
&& \qquad \qquad + \frac{g_\MSbar^4}{(16 \pi^2)^2} \Bigg[ N_c^2 \Bigg(-\frac{29}{30} + \frac{1633}{1440} \alpha_\MSbar - \frac{9}{40} \alpha_\MSbar^2 -\frac{1}{4} \alpha_\MSbar^3 + \nonumber \\
&& \qquad \qquad \qquad \qquad \qquad \ \ \ \left(\frac{23}{18} + \frac{5}{6} \alpha_\MSbar - \frac{5}{8} \alpha_\MSbar^2 \right) \logq + \nonumber \\
&& \qquad \qquad \qquad \qquad \qquad \ \ \ \left(\frac{7}{24} - \frac{1}{4} \alpha_\MSbar \right) \logqsq + \nonumber \\
&& \qquad \qquad \qquad \qquad \qquad \ \ \ \left(-\frac{47}{20} - \frac{3}{5} \alpha_\MSbar + \frac{1}{20} \alpha_\MSbar^2 \right) \zeta (3) \Bigg) + \nonumber \\
&& \qquad \qquad \qquad \quad \ \ \ \ \ \frac{N_f}{N_c} \Bigg(\frac{2}{9} + \frac{4}{9} \logq \Bigg) + \nonumber \\
&& \qquad \qquad \qquad \ \ \ \ \ N_f N_c \Bigg(\frac{157}{108} -\frac{11}{9} \logq + 2 \zeta (3) \Bigg) \Bigg] + \mathcal{O} (g_\MSbar^6) \Bigg\} + \nonumber \\
&& \qquad G_4 \Bigg\{\frac{g_\MSbar^2}{16 \pi^2} N_c \ \left(-\frac{4}{3} + 2 \alpha_\MSbar \right) \qquad \qquad \qquad \qquad \qquad \qquad \qquad \qquad \qquad \qquad \qquad \qquad \nonumber \\
&& \qquad \qquad + \frac{g_\MSbar^4}{(16 \pi^2)^2} \Bigg[ N_c^2 \Bigg(\frac{611}{120} + \frac{1183}{360} \alpha_\MSbar - \frac{7}{5} \alpha_\MSbar^2 - \alpha_\MSbar^3 + \nonumber \\
&& \qquad \qquad \qquad \qquad \qquad \ \ \ \left(\frac{7}{2} + \frac{5}{2} \alpha_\MSbar - 2 \alpha_\MSbar^2 \right) \logq + \nonumber \\
&& \qquad \qquad \qquad \qquad \qquad \ \ \ \left(-\frac{99}{10} - \frac{19}{10} \alpha_\MSbar + \frac{1}{5} \alpha_\MSbar^2 \right) \zeta (3) \Bigg) + \nonumber \\
&& \qquad \qquad \qquad \quad \ \ \ \ \ \frac{N_f}{N_c} \Bigg(\frac{8}{9} + \frac{16}{9} \logq \Bigg) + \nonumber \\
&& \qquad \qquad \qquad \ \ \ \ \ N_f N_c \Bigg(-\frac{22}{9} - \frac{8}{3} \logq  + 8 \zeta (3) \Bigg) \Bigg] + \mathcal{O} (g_\MSbar^6) \Bigg\},
\label{MSbar_first}
\end{eqnarray}
\begin{eqnarray}
&& G^\MSbar_{g2} (q,-q) = \qquad \qquad \qquad \qquad \qquad \qquad \qquad \qquad \qquad \qquad \qquad \qquad\qquad \qquad \qquad \qquad \quad \nonumber \\
&& \qquad G_1 \Bigg\{\frac{g_\MSbar^2}{16 \pi^2} N_f \ \left(\frac{4}{9} - \frac{2}{3} \logq \right) \nonumber \\
&& \qquad \qquad + \frac{g_\MSbar^4}{(16 \pi^2)^2} \Bigg[ \frac{N_f}{N_c} \Bigg(-\frac{425}{162} + \frac{7}{9} \logq + \frac{4}{9} \logqsq + 4 \zeta (3) \Bigg) + \nonumber \\
&& \qquad \qquad \qquad \ \ \ \ \ N_f N_c \Bigg(\frac{179}{180} - \frac{29}{36} \alpha_\MSbar - \frac{2}{9} \alpha_\MSbar^2 + \nonumber \\
&& \qquad \qquad \qquad \qquad \qquad \ \ \ \ \left(- \frac{11}{9} + \frac{10}{9} \alpha_\MSbar + \frac{1}{3} \alpha_\MSbar^2 \right) \logq + \nonumber \\
&& \qquad \qquad \qquad \qquad \qquad \ \ \ \ \left(- \frac{2}{3} + \frac{1}{3} \alpha_\MSbar \right) \logqsq + \frac{28}{5} \zeta (3) \Bigg) \Bigg] + \mathcal{O} (g_\MSbar^6) \Bigg\} + \nonumber \\
&& \qquad G_2 \Bigg\{\frac{g_\MSbar^2}{16 \pi^2} N_f \ \left(- \frac{1}{3} - \frac{1}{3} \alpha_\MSbar \right) \nonumber \\
&& \qquad \qquad + \frac{g_\MSbar^4}{(16 \pi^2)^2} \Bigg[ \frac{N_f}{N_c} \Bigg(\frac{13}{18} + \frac{13}{18} \alpha_\MSbar + \left( \frac{4}{9} + \frac{4}{9} \alpha_\MSbar \right) \logq \Bigg) + \nonumber \\
&& \qquad \qquad \qquad \ \ \ \ \ \ N_f N_c \Bigg(-\frac{457}{540} - \frac{133}{60} \alpha_\MSbar + \frac{5}{18} \alpha_\MSbar^2  + \frac{1}{6} \alpha_\MSbar^3 + \nonumber \\
&& \qquad \qquad \qquad \qquad \qquad \ \ \ \ \left(- \frac{17}{18} + \frac{1}{3} \alpha_\MSbar + \frac{1}{2} \alpha_\MSbar^2 \right) \logq + \nonumber \\
&& \qquad \qquad \qquad \qquad \qquad \ \ \ \ \frac{1}{6} \logqsq + \left(\frac{12}{5} + \frac{12}{5} \alpha_\MSbar \right) \zeta (3) \Bigg) \Bigg] + \mathcal{O} (g_\MSbar^6) \Bigg\} + \nonumber 
\end{eqnarray}
\begin{eqnarray}
&& \qquad G_3 \Bigg\{\frac{g_\MSbar^2}{16 \pi^2} N_f \ \frac{1}{3} + \frac{g_\MSbar^4}{(16 \pi^2)^2} \Bigg[ \frac{N_f}{N_c} \Bigg(-\frac{13}{18} - \frac{4}{9} \logq \Bigg) + \qquad \qquad \qquad \qquad \qquad \qquad  \nonumber \\
&& \qquad \qquad \qquad \qquad \qquad \quad \ \ \ \ \ N_f N_c \Bigg(\frac{457}{540} - \frac{1}{9} \alpha_\MSbar - \frac{1}{6} \alpha_\MSbar^2 + \left(\frac{17}{18} - \frac{1}{2} \alpha_\MSbar \right) \logq - \nonumber \\
&& \qquad \qquad \qquad \qquad \qquad \qquad \qquad \quad \ \ \ \ \frac{1}{6} \logqsq - \frac{12}{5} \zeta (3) \Bigg) \Bigg] + \mathcal{O} (g_\MSbar^6) \Bigg\} + \nonumber \\
&& \qquad G_4 \Bigg\{\frac{g_\MSbar^2}{16 \pi^2} N_f \ \frac{4}{3} + \frac{g_\MSbar^4}{(16 \pi^2)^2} \Bigg[ \frac{N_f}{N_c} \Bigg(-\frac{26}{9} - \frac{16}{9} \logq \Bigg) + \nonumber \\
&& \qquad \qquad \qquad \qquad \qquad \quad \ \ \ \ \ N_f N_c \Bigg(\frac{419}{45} - \frac{4}{9} \alpha_\MSbar - \frac{2}{3} \alpha_\MSbar^2 + \left(\frac{2}{3} - 2 \alpha_\MSbar \right) \logq - \nonumber \\
&& \qquad \qquad \qquad \qquad \qquad \qquad \qquad \quad \ \ \ \ \frac{48}{5} \zeta (3) \Bigg) \Bigg] + \mathcal{O} (g_\MSbar^6) \Bigg\}, 
\end{eqnarray}
\begin{eqnarray}
&& G^\MSbar_{g3} (q,-q) = G^\MSbar_{g4} (q,-q) = \qquad \qquad \qquad \qquad \qquad \qquad \qquad \qquad \qquad \qquad \qquad \qquad\qquad \qquad \qquad \quad \nonumber \\
&& \qquad G_2 \Bigg\{ 1 + \frac{g_\MSbar^2}{16 \pi^2} N_c \ \left(\frac{1}{4} + \alpha_\MSbar + \frac{1}{4} \alpha_\MSbar^2 + \frac{1}{2} \logq \right) \nonumber \\
&& \qquad \qquad + \frac{g_\MSbar^4}{(16 \pi^2)^2} \Bigg[ N_c^2 \Bigg(\frac{569}{120} + \frac{5713}{720} \alpha_\MSbar + \frac{269}{1440} \alpha_\MSbar^2 - \frac{77}{120} \alpha_\MSbar^3 - \frac{1}{16} \alpha_\MSbar^4  + \nonumber \\
&& \qquad \qquad \qquad \qquad \qquad \ \ \left(- \frac{25}{72} - \frac{5}{8} \alpha_\MSbar - \frac{11}{24} \alpha_\MSbar^2 - \frac{1}{4} \alpha_\MSbar^3 \right) \logq + \nonumber \\
&& \qquad \qquad \qquad \qquad \qquad \ \ \left(\frac{7}{24} - \frac{1}{4} \alpha_\MSbar \right) \logqsq + \nonumber \\
&& \qquad \qquad \qquad \qquad \qquad \ \ \left(- \frac{47}{20} - \frac{123}{40} \alpha_\MSbar - \frac{17}{40} \alpha_\MSbar^2 + \frac{1}{20} \alpha_\MSbar^3 \right) \zeta (3) \Bigg) + \nonumber \\
&& \qquad \qquad \qquad \quad \ \ N_f N_c \Bigg(- \frac{17}{60} - \frac{49}{45} \alpha_\MSbar - \frac{2}{9} \alpha_\MSbar^2 + \nonumber \\
&& \qquad \qquad \qquad \qquad \qquad \ \ \ \left( \frac{2}{9} - \frac{1}{6} \alpha_\MSbar - \frac{1}{6} \alpha_\MSbar^2 \right) \logq  - \frac{1}{6} \logqsq + \nonumber \\
&& \qquad \qquad \qquad \qquad \qquad \ \ \ \left(- \frac{2}{5} - \frac{2}{5} \alpha_\MSbar \right) \zeta (3) \Bigg) \Bigg] + \mathcal{O} (g_\MSbar^6) \Bigg\} + \nonumber \\
&& \qquad G_1 \Bigg\{\frac{g_\MSbar^2}{16 \pi^2} N_c \ \left(\frac{7}{4} + \alpha_\MSbar + \frac{1}{4} \alpha_\MSbar^2 \right) \qquad \qquad \qquad \qquad \qquad \qquad \qquad \qquad \qquad \qquad \qquad \nonumber \\
&& \qquad \qquad + \frac{g_\MSbar^4}{(16 \pi^2)^2} \Bigg[ N_c^2 \Bigg(\frac{866}{45} + \frac{2071}{720} \alpha_\MSbar - \frac{779}{360} \alpha_\MSbar^2 - \frac{5}{16} \alpha_\MSbar^3 - \frac{1}{8} \alpha_\MSbar^4  + \nonumber \\
&& \qquad \qquad \qquad \qquad \quad \ \ \ \ \ \ \left(-\frac{7}{4} + \frac{7}{24} \alpha_\MSbar - \frac{1}{6} \alpha_\MSbar^2 - \frac{3}{8} \alpha_\MSbar^3 \right) \logq + \nonumber \\
&& \qquad \qquad \qquad \qquad \quad \ \ \ \ \ \ \left(- \frac{69}{10} - \frac{7}{5} \alpha_\MSbar + \frac{1}{5} \alpha_\MSbar^2 \right) \zeta (3) \Bigg) + \nonumber \\
&& \qquad \qquad \qquad \ \ \ \ \ N_f N_c \Bigg(-\frac{599}{180} + \frac{1}{36} \alpha_\MSbar + \frac{5}{9} \alpha_\MSbar^2 + \nonumber \\
&& \qquad \qquad \qquad \qquad \quad \ \ \ \ \ \ \left(- \frac{2}{3} \alpha_\MSbar - \frac{1}{3} \alpha_\MSbar^2 \right) \logq + \frac{2}{5} \zeta (3) \Bigg) \Bigg] + \mathcal{O} (g_\MSbar^6) \Bigg\} + \nonumber
\end{eqnarray}
\begin{eqnarray}
&& G_3 \Bigg\{\frac{g_\MSbar^2}{16 \pi^2} N_c \ \left(- \frac{1}{4} - \frac{1}{4} \alpha_\MSbar - \frac{1}{2} \logq \right) \nonumber \\
&& \qquad \ + \frac{g_\MSbar^4}{(16 \pi^2)^2} \Bigg[ N_c^2 \Bigg(-\frac{569}{120} - \frac{1373}{1440} \alpha_\MSbar + \frac{109}{240} \alpha_\MSbar^2 +\frac{1}{16} \alpha_\MSbar^3 + \nonumber \\
&& \qquad \qquad \qquad \quad \ \ \ \ \ \ \left(\frac{25}{72} - \frac{1}{8} \alpha_\MSbar + \frac{3}{8} \alpha_\MSbar^2 \right) \logq + \nonumber \\
&& \qquad \qquad \qquad \quad \ \ \ \ \ \ \left(-\frac{7}{24} + \frac{1}{4} \alpha_\MSbar \right) \logqsq + \nonumber \\
&& \qquad \qquad \qquad \quad \ \ \ \ \ \ \left(\frac{47}{20} + \frac{3}{5} \alpha_\MSbar - \frac{1}{20} \alpha_\MSbar^2 \right) \zeta (3) \Bigg) + \nonumber \\
&& \qquad \qquad \ \ \ \ \ \ N_f N_c \Bigg(\frac{17}{60} + \frac{2}{9} \alpha_\MSbar + \left(-\frac{2}{9} + \frac{1}{6} \alpha_\MSbar \right) \logq + \nonumber \\
&& \qquad \qquad \qquad \quad \ \ \ \ \ \ \frac{1}{6} \logqsq + \frac{2}{5} \zeta (3) \Bigg) \Bigg] + \mathcal{O} (g_\MSbar^6) \Bigg\} + \nonumber \\
&& G_4 \Bigg\{\frac{g_\MSbar^2}{16 \pi^2} N_c \ \left(-3 - \alpha_\MSbar \right) \nonumber \\
&& \qquad \ + \frac{g_\MSbar^4}{(16 \pi^2)^2} \Bigg[ N_c^2 \Bigg(-\frac{1117}{40} - \frac{923}{360} \alpha_\MSbar + \frac{139}{60} \alpha_\MSbar^2 + \frac{1}{4} \alpha_\MSbar^3 + \nonumber \\
&& \qquad \qquad \qquad \quad \ \ \ \ \ \ \left(3 + \frac{1}{3} \alpha_\MSbar + \alpha_\MSbar^2 \right) \logq + \nonumber \\
&& \qquad \qquad \qquad \quad \ \ \ \ \ \ \left(\frac{99}{10} + \frac{19}{10} \alpha_\MSbar - \frac{1}{5} \alpha_\MSbar^2 \right) \zeta (3) \Bigg) + \nonumber \\
&& \qquad \qquad \ \ \ \ \ \ N_f N_c \Bigg(\frac{52}{15} + \frac{8}{9} \alpha_\MSbar + \frac{2}{3} \alpha_\MSbar \logq  + \frac{8}{5} \zeta (3) \Bigg) \Bigg] + \mathcal{O} (g_\MSbar^6) \Bigg\}, 
\end{eqnarray}
\begin{eqnarray}
&& G^\MSbar_{g5} (q,-q) = \qquad \qquad \qquad \qquad \qquad \qquad \qquad \qquad \qquad \qquad \qquad \qquad\qquad \qquad \qquad \qquad \nonumber \\
&& \qquad G_3 \Bigg\{ 1 + \mathcal{O} (g_\MSbar^6) \Bigg\} + \nonumber \\
&& \qquad \left( G_2 - G_3 \right) \Bigg\{\frac{g_\MSbar^2}{16 \pi^2} \Bigg[ N_c \ \left(\frac{97}{36} + \frac{1}{2} \alpha_\MSbar + \frac{1}{4} \alpha_\MSbar^2 + \left(-\frac{13}{6} + \frac{1}{2} \alpha_\MSbar \right) \logq \right) +  \nonumber \\
&& \qquad \qquad \qquad \qquad \qquad \ N_f \ \left(-\frac{10}{9} + \frac{2}{3} \logq \right) \Bigg] \nonumber \\ 
&& \qquad \qquad \qquad \quad \ + \frac{g_\MSbar^4}{(16 \pi^2)^2} \Bigg[ N_c^2 \Bigg(\frac{2381}{96} - \frac{463}{288} \alpha_\MSbar - \frac{95}{144} \alpha_\MSbar^2 + \frac{1}{16} \alpha_\MSbar^3 - \frac{1}{16} \alpha_\MSbar^4  + \nonumber \\
&& \qquad \qquad \qquad \qquad \qquad \qquad \qquad \left(-\frac{137}{12} + \frac{13}{36} \alpha_\MSbar + \frac{11}{24} \alpha_\MSbar^2 - \frac{3}{8} \alpha_\MSbar^3 \right) \logq + \nonumber \\
&& \qquad \qquad \qquad \qquad \qquad \qquad \qquad \left(\frac{13}{8} + \frac{17}{24} \alpha_\MSbar - \frac{1}{4} \alpha_\MSbar^2 \right) \logqsq + \nonumber \\
&& \qquad \qquad \qquad \qquad \qquad \qquad \qquad \ \left(-3 + 2 \alpha_\MSbar \right) \zeta (3) \Bigg) + \nonumber \\
&& \qquad \qquad \qquad \quad \qquad \qquad \quad \frac{N_f}{N_c} \Bigg(\frac{55}{12} - \logq - 4 \zeta (3) \Bigg) + \nonumber \\
&& \qquad \qquad \qquad \quad \qquad \qquad N_f N_c \Bigg(-\frac{287}{24} + \frac{5}{9} \alpha_\MSbar + \frac{5}{9} \alpha_\MSbar^2 + \nonumber \\
&& \qquad \qquad \qquad \qquad \qquad \qquad \qquad \left(\frac{31}{6} + \frac{2}{9} \alpha_\MSbar - \frac{1}{3} \alpha_\MSbar^2 \right) \logq + \nonumber \\
&& \qquad \qquad \qquad \qquad \qquad \qquad \qquad \left(-\frac{1}{2} - \frac{1}{3} \alpha_\MSbar \right) \logqsq \Bigg) \Bigg] + \mathcal{O} (g_\MSbar^6) \Bigg\}.
\end{eqnarray}

\bigskip

The expressions for the Green's functions with two external gluon fields and nonzero-momentum operator insertion ($G^\MSbar_{gi} (q,0)$) are as follows:
\begin{eqnarray}
&& G^\MSbar_{g1} (q,0) = \qquad \qquad \qquad \qquad \qquad \qquad \qquad \qquad \qquad \qquad \qquad \qquad\qquad \qquad \qquad \qquad \nonumber \\
&& \qquad G_1 \Bigg\{\frac{g_\MSbar^2}{16 \pi^2} N_c \ \frac{1}{2} + \frac{g_\MSbar^4}{(16 \pi^2)^2} \Bigg[ N_c^2 \Bigg(\frac{737}{720} + \frac{161}{480} \alpha_\MSbar - \frac{1}{8} \alpha_\MSbar^2 + \nonumber \\
&& \qquad \qquad \qquad \qquad \qquad \qquad \qquad \quad \ \ \left(- \frac{7}{8} - \frac{3}{8} \alpha_\MSbar \right) \logq + \left( \frac{13}{40} - \frac{1}{10} \alpha_\MSbar \right) \zeta (3) \Bigg) + \nonumber \\
&& \qquad \qquad \qquad \qquad \qquad \qquad \ \ N_f N_c \Bigg(- \frac{8}{9} + \frac{1}{3} \logq \Bigg) \Bigg] + \mathcal{O} (g_\MSbar^6) \Bigg\} + \nonumber
\end{eqnarray}
\begin{eqnarray}
&& \qquad \left(G_5 - G_6 \right) \Bigg\{\frac{g_\MSbar^2}{16 \pi^2} N_c \ \left(\frac{1}{2} - \frac{1}{2} \logq \right) \qquad \qquad \qquad \qquad \qquad \qquad \qquad \qquad \qquad \qquad \nonumber \\
&& \qquad \qquad \quad \qquad \ + \frac{g_\MSbar^4}{(16 \pi^2)^2} \Bigg[ N_c^2 \Bigg(\frac{1037}{240} + \frac{11}{160} \alpha_\MSbar - \frac{1}{8} \alpha_\MSbar^2 + \nonumber \\
&& \qquad \qquad \qquad \qquad \qquad \qquad \qquad \ \left(- \frac{5}{18} - \frac{1}{3} \alpha_\MSbar + \frac{1}{8} \alpha_\MSbar^2 \right) \logq + \nonumber \\
&& \qquad \qquad \qquad \qquad \qquad \qquad \qquad \ \left(- \frac{7}{24} + \frac{1}{4} \alpha_\MSbar \right) \logqsq + \nonumber \\
&& \qquad \qquad \qquad \qquad \qquad \qquad \qquad \ \left( - \frac{43}{20} + \frac{1}{5} \alpha_\MSbar \right) \zeta (3) \Bigg) + \nonumber \\
&& \qquad \qquad \qquad \qquad \quad \qquad \ N_f N_c \Bigg(-\frac{187}{108} + \frac{5}{9} \logq \Bigg) \Bigg] + \mathcal{O} (g_\MSbar^6) \Bigg\} + \nonumber \\
&& \qquad G_7 \Bigg\{\frac{g_\MSbar^4}{{(16 \pi^2)}^2} N_c^2 \Bigg(\frac{83}{160} \alpha_\MSbar + \frac{31}{160} \alpha_\MSbar^2 + \left( \frac{1}{8} \alpha_\MSbar - \frac{1}{8} \alpha_\MSbar^2 \right) \logq + \nonumber \\
&& \qquad \qquad \qquad \quad \qquad \ \ \left(- \frac{1}{40} \alpha_\MSbar + \frac{3}{40} \alpha_\MSbar^2 \right) \zeta (3) \Bigg) + \mathcal{O} (g_\MSbar^6) \Bigg\} + \nonumber \\
&& \qquad G_4 \Bigg\{\frac{g_\MSbar^4}{{(16 \pi^2)}^2} N_c^2 \Bigg(- \frac{83}{80} - \frac{31}{80} \alpha_\MSbar + \left(- \frac{1}{4} + \frac{1}{4} \alpha_\MSbar \right) \logq + \nonumber \\
&& \qquad \qquad \qquad \quad \qquad \ \ \left(\frac{1}{20} - \frac{3}{20} \alpha_\MSbar \right) \zeta (3) \Bigg) + \mathcal{O} (g_\MSbar^6) \Bigg\},
\end{eqnarray}
\begin{eqnarray}
&& G^\MSbar_{g2} (q,0) = \qquad \qquad \qquad \qquad \qquad \qquad \qquad \qquad \qquad \qquad \qquad \qquad\qquad \qquad \qquad \qquad \nonumber \\
&& \qquad G_1 \Bigg\{\frac{g_\MSbar^4}{(16 \pi^2)^2} N_c N_f \Bigg(\frac{91}{90} - \frac{1}{3} \logq - \frac{1}{5} \zeta (3) \Bigg) + \mathcal{O} (g_\MSbar^6) \Bigg\} + \nonumber \\
&& \qquad \left(G_5 - G_6 \right) \Bigg\{\frac{g_\MSbar^4}{(16 \pi^2)^2} N_c N_f \Bigg(\frac{683}{540} - \frac{7}{9} \logq + \frac{1}{6} \logqsq + \frac{2}{5} \zeta (3) \Bigg) \nonumber \\
&& \qquad \qquad \qquad \quad \ + \mathcal{O} (g_\MSbar^6) \Bigg\} + \nonumber \\
&& \qquad G_7 \Bigg\{\frac{g_\MSbar^4}{{(16 \pi^2)}^2} N_c N_f \Bigg(- \frac{3}{10} \alpha_\MSbar + \frac{2}{5} \alpha_\MSbar \zeta (3) \Bigg) + \mathcal{O} (g_\MSbar^6) \Bigg\} + \nonumber \\
&& \qquad G_4 \Bigg\{\frac{g_\MSbar^4}{{(16 \pi^2)}^2} N_c N_f \Bigg(\frac{3}{5} - \frac{4}{5} \zeta (3) \Bigg) + \mathcal{O} (g_\MSbar^6) \Bigg\},
\end{eqnarray}
\begin{eqnarray}
&& G^\MSbar_{g3} (q,0) = \qquad \qquad \qquad \qquad \qquad \qquad \qquad \qquad \qquad \qquad \qquad \qquad\qquad \qquad \qquad \qquad \nonumber \\
&& \qquad G_1 \Bigg\{\frac{g_\MSbar^2}{16 \pi^2} N_c \ \left(- \frac{1}{4} - \frac{1}{4} \alpha_\MSbar \right) \nonumber \\
&& \qquad \qquad + \frac{g_\MSbar^4}{(16 \pi^2)^2} \Bigg[ N_c^2 \Bigg(- \frac{187}{360} - \frac{707}{1440} \alpha_\MSbar - \frac{1}{24} \alpha_\MSbar^2 + \frac{1}{16} \alpha_\MSbar^3 + \nonumber \\
&& \qquad \qquad \qquad \qquad \qquad \ \ \ \left(\frac{1}{2} + \frac{1}{12} \alpha_\MSbar + \frac{1}{4} \alpha_\MSbar^2 \right) \logq + \nonumber \\
&& \qquad \qquad \qquad \qquad \qquad \ \ \ \left(- \frac{9}{10} - \frac{9}{40} \alpha_\MSbar - \frac{1}{4} \alpha_\MSbar^2 \right) \zeta (3) \Bigg) + \nonumber \\
&& \qquad \qquad \qquad \ \ \ \ \ N_f N_c \Bigg(- \frac{1}{18} - \frac{5}{18} \alpha_\MSbar + \frac{1}{6} \alpha_\MSbar \logq \Bigg) \Bigg] + \mathcal{O} (g_\MSbar^6) \Bigg\} + \nonumber \\
&& \qquad \left(G_5 - G_6 \right) \Bigg\{\frac{g_\MSbar^2}{16 \pi^2} N_c \ \left(- \frac{5}{4} - \frac{1}{4} \alpha_\MSbar + \frac{1}{2} \logq \right) \nonumber \\
&& \qquad \qquad \quad \qquad \ + \frac{g_\MSbar^4}{(16 \pi^2)^2} \Bigg[ N_c^2 \Bigg(-\frac{1459}{240} - \frac{971}{1440} \alpha_\MSbar + \frac{7}{48} \alpha_\MSbar^2 + \frac{1}{16} \alpha_\MSbar^3 + \nonumber \\
&& \qquad \qquad \qquad \qquad \qquad \qquad \ \ \ \ \ \ \ \left( \frac{101}{72} + \frac{13}{24} \alpha_\MSbar + \frac{1}{8} \alpha_\MSbar^2 \right) \logq + \nonumber \\
&& \qquad \qquad \qquad \qquad \qquad \qquad \ \ \ \ \ \ \ \left( \frac{7}{24} - \frac{1}{4} \alpha_\MSbar \right) \logqsq + \nonumber \\
&& \qquad \qquad \qquad \qquad \qquad \qquad \ \ \ \ \ \ \ \left( \frac{3}{10} - \frac{3}{10} \alpha_\MSbar - \frac{1}{4} \alpha_\MSbar^2 \right) \zeta (3) \Bigg) + \nonumber \\
&& \qquad \qquad \qquad \qquad \quad \ \ \ \ \ \ N_f N_c \Bigg( \frac{7}{12} - \frac{5}{18} \alpha_\MSbar + \left(\frac{2}{9} + \frac{1}{6} \alpha_\MSbar \right) \logq - \nonumber \\
&& \qquad \qquad \qquad \qquad \qquad \qquad \ \ \ \ \ \ \ \ \frac{1}{6} \logqsq \Bigg) \Bigg] + \mathcal{O} (g_\MSbar^6) \Bigg\} + \nonumber \\
&& \qquad G_7 \Bigg\{\frac{g_\MSbar^2}{16 \pi^2} N_c \ \left(\frac{1}{4} \alpha_\MSbar - \frac{1}{4} \alpha_\MSbar^2 \right) \qquad \qquad \qquad \qquad \qquad \qquad \qquad \qquad \qquad \qquad \qquad \nonumber \\ 
&& \qquad \quad \ \ +\frac{g_\MSbar^4}{{(16 \pi^2)}^2} \Bigg[ N_c^2 \Bigg(\frac{487}{480} \alpha_\MSbar - \frac{311}{1440} \alpha_\MSbar^2 - \frac{1}{16} \alpha_\MSbar^3 + \frac{1}{16} \alpha_\MSbar^4 + \nonumber \\
&& \qquad \qquad \qquad \qquad \qquad \left(- \frac{1}{2} \alpha_\MSbar - \frac{1}{6} \alpha_\MSbar^2 + \frac{1}{4} \alpha_\MSbar^3 \right) \logq + \nonumber \\
&& \qquad \qquad \qquad \qquad \qquad \left( \frac{7}{40} \alpha_\MSbar - \frac{4}{5} \alpha_\MSbar^2 - \frac{3}{8} \alpha_\MSbar^3 \right) \zeta (3) \Bigg) + \nonumber \\
&& \qquad \qquad \qquad \quad N_f N_c \Bigg(- \frac{1}{6} \alpha_\MSbar - \frac{5}{18} \alpha_\MSbar^2 + \frac{1}{6} \alpha_\MSbar^2 \logq \Bigg) \Bigg] + \mathcal{O} (g_\MSbar^6) \Bigg\} + \nonumber
\end{eqnarray}
\begin{eqnarray}
&& G_4 \Bigg\{\frac{g_\MSbar^2}{16 \pi^2} N_c \ \left(-\frac{1}{2} + \frac{1}{2} \alpha_\MSbar \right) \nonumber \\
&& \qquad \quad \ \ +\frac{g_\MSbar^4}{{(16 \pi^2)}^2} \Bigg[ N_c^2 \Bigg(- \frac{487}{240} + \frac{311}{720} \alpha_\MSbar + \frac{1}{8} \alpha_\MSbar^2 - \frac{1}{8} \alpha_\MSbar^3 + \nonumber \\
&& \qquad \qquad \qquad \qquad \qquad \left(1 + \frac{1}{3} \alpha_\MSbar - \frac{1}{2} \alpha_\MSbar^2 \right) \logq + \nonumber \\
&& \qquad \qquad \qquad \qquad \qquad \left(- \frac{7}{20} + \frac{8}{5} \alpha_\MSbar + \frac{3}{4} \alpha_\MSbar^2 \right) \zeta (3) \Bigg) + \nonumber \\
&& \qquad \qquad \qquad \quad N_f N_c \Bigg(\frac{1}{3} + \frac{5}{9} \alpha_\MSbar - \frac{1}{3} \alpha_\MSbar \logq \Bigg) \Bigg] + \mathcal{O} (g_\MSbar^6) \Bigg\}, \qquad \qquad
\end{eqnarray}
\begin{eqnarray}
&& G^\MSbar_{g4} (q,0) = \qquad \qquad \qquad \qquad \qquad \qquad \qquad \qquad \qquad \qquad \qquad \qquad\qquad \qquad \qquad \qquad \nonumber \\
&& \qquad G_5 \Bigg\{1 + \mathcal{O} (g_\MSbar^6) \Bigg\} + \nonumber \\
&& \qquad \left(G_5 - G_6 \right) \Bigg\{\frac{g_\MSbar^2}{16 \pi^2} N_c \ \left(- \frac{3}{4} + \frac{1}{4} \alpha_\MSbar + \frac{1}{2} \logq \right) \nonumber \\
&& \qquad \qquad \quad \qquad \ + \frac{g_\MSbar^4}{(16 \pi^2)^2} \Bigg[ N_c^2 \Bigg(-\frac{1199}{240} - \frac{391}{1440} \alpha_\MSbar + \frac{13}{48} \alpha_\MSbar^2 - \frac{1}{16} \alpha_\MSbar^3 + \nonumber \\
&& \qquad \qquad \qquad \qquad \qquad \qquad \qquad \ \left( \frac{47}{72} + \frac{5}{8} \alpha_\MSbar - \frac{3}{8} \alpha_\MSbar^2 \right) \logq + \nonumber \\
&& \qquad \qquad \qquad \qquad \qquad \qquad \qquad \ \left( \frac{7}{24} - \frac{1}{4} \alpha_\MSbar \right) \logqsq + \nonumber \\
&& \qquad \qquad \qquad \qquad \qquad \qquad \qquad \ \left( \frac{9}{5} + \frac{3}{40} \alpha_\MSbar + \frac{1}{8} \alpha_\MSbar^2 \right) \zeta (3) \Bigg) + \nonumber \\
&& \qquad \qquad \qquad \qquad \quad \qquad N_f N_c \Bigg( \frac{1}{2} + \frac{5}{18} \alpha_\MSbar + \left(\frac{2}{9} - \frac{1}{6} \alpha_\MSbar \right) \logq - \nonumber \\
&& \qquad \qquad \qquad \qquad \qquad \qquad \qquad \ \frac{1}{6} \logqsq \Bigg) \Bigg] + \mathcal{O} (g_\MSbar^6) \Bigg\} + \nonumber
\end{eqnarray}
\begin{eqnarray}
&& \qquad G_1 \Bigg\{\frac{g_\MSbar^2}{16 \pi^2} N_c \ \left(\frac{1}{4} + \frac{1}{4} \alpha_\MSbar \right) \nonumber \\
&& \qquad \qquad + \frac{g_\MSbar^4}{(16 \pi^2)^2} \Bigg[ N_c^2 \Bigg(\frac{203}{360} - \frac{127}{1440} \alpha_\MSbar + \frac{1}{12} \alpha_\MSbar^2 - \frac{1}{16} \alpha_\MSbar^3 + \nonumber \\
&& \qquad \qquad \qquad \qquad \qquad \quad \left(- \frac{1}{4} + \frac{1}{6} \alpha_\MSbar - \frac{1}{4} \alpha_\MSbar^2 \right) \logq + \nonumber \\
&& \qquad \qquad \qquad \qquad \qquad \quad \left(\frac{3}{5} + \frac{3}{20} \alpha_\MSbar + \frac{1}{8} \alpha_\MSbar^2 \right) \zeta (3) \Bigg) + \nonumber \\
&& \qquad \qquad \qquad \quad \ \ N_f N_c \Bigg(- \frac{5}{36} + \frac{5}{18} \alpha_\MSbar - \frac{1}{6} \alpha_\MSbar \logq \Bigg) \Bigg] + \mathcal{O} (g_\MSbar^6) \Bigg\} + \nonumber \\
&& \qquad G_7 \Bigg\{\frac{g_\MSbar^2}{16 \pi^2} N_c \ \left(- \frac{1}{4} \alpha_\MSbar + \frac{1}{4} \alpha_\MSbar^2 \right) \qquad \qquad \qquad \qquad \qquad \qquad \qquad \qquad \qquad \qquad \quad \nonumber \\ 
&& \qquad \quad \ \ \ \ +\frac{g_\MSbar^4}{{(16 \pi^2)}^2} \Bigg[ N_c^2 \Bigg(- \frac{191}{160} \alpha_\MSbar - \frac{571}{1440} \alpha_\MSbar^2 + \frac{7}{48} \alpha_\MSbar^3 - \frac{1}{16} \alpha_\MSbar^4 + \nonumber \\
&& \qquad \qquad \qquad \qquad \qquad \ \ \ \left(\frac{1}{4} \alpha_\MSbar + \frac{5}{12} \alpha_\MSbar^2 - \frac{1}{4} \alpha_\MSbar^3 \right) \logq + \nonumber \\
&& \qquad \qquad \qquad \qquad \qquad \ \ \ \left(- \frac{13}{40} \alpha_\MSbar + \frac{1}{5} \alpha_\MSbar^2 + \frac{1}{8} \alpha_\MSbar^3 \right) \zeta (3) \Bigg) + \nonumber \\
&& \qquad \qquad \qquad \quad \ \ \ N_f N_c \Bigg(\frac{1}{3} \alpha_\MSbar + \frac{5}{18} \alpha_\MSbar^2 - \frac{1}{6} \alpha_\MSbar^2 \logq \Bigg) \Bigg] + \mathcal{O} (g_\MSbar^6) \Bigg\} + \nonumber \\
&& \qquad G_4 \Bigg\{\frac{g_\MSbar^2}{16 \pi^2} N_c \ \left(\frac{1}{2} - \frac{1}{2} \alpha_\MSbar \right) \nonumber \\
&& \qquad \quad \ \ \ +\frac{g_\MSbar^4}{{(16 \pi^2)}^2} \Bigg[ N_c^2 \Bigg(\frac{191}{80} + \frac{571}{120} \alpha_\MSbar - \frac{7}{24} \alpha_\MSbar^2 + \frac{1}{8} \alpha_\MSbar^3 + \nonumber \\
&& \qquad \qquad \qquad \qquad \qquad \ \ \ \left(-\frac{1}{2} - \frac{5}{6} \alpha_\MSbar + \frac{1}{2} \alpha_\MSbar^2 \right) \logq + \nonumber \\
&& \qquad \qquad \qquad \qquad \qquad \ \ \ \left(\frac{13}{20} - \frac{2}{5} \alpha_\MSbar - \frac{1}{4} \alpha_\MSbar^2 \right) \zeta (3) \Bigg) + \nonumber \\
&& \qquad \qquad \qquad \quad \ \ N_f N_c \Bigg(- \frac{2}{3} - \frac{5}{9} \alpha_\MSbar + \frac{1}{3} \alpha_\MSbar \logq \Bigg) \Bigg] + \mathcal{O} (g_\MSbar^6) \Bigg\},
\end{eqnarray}
\begin{eqnarray}
&& G^\MSbar_{g5} (q,0) = \qquad \qquad \qquad \qquad \qquad \qquad \qquad \qquad \qquad \qquad \qquad \qquad\qquad \qquad \qquad \qquad \nonumber \\
&& \qquad G_6 \Bigg\{ 1 + \mathcal{O} (g_\MSbar^6) \Bigg\} + \nonumber \\
&& \qquad \left( G_5 - G_6 \right) \Bigg\{\frac{g_\MSbar^2}{16 \pi^2} \Bigg[ N_c \ \left(\frac{97}{36} + \frac{1}{2} \alpha_\MSbar + \frac{1}{4} \alpha_\MSbar^2 + \left(-\frac{13}{6} + \frac{1}{2} \alpha_\MSbar \right) \logq \right) +  \nonumber \\
&& \qquad \qquad \qquad \qquad \qquad \ N_f \ \left(-\frac{10}{9} + \frac{2}{3} \logq \right) \Bigg] \nonumber \\ 
&& \qquad \qquad \qquad \quad \ + \frac{g_\MSbar^4}{(16 \pi^2)^2} \Bigg[ N_c^2 \Bigg(\frac{2381}{96} - \frac{463}{288} \alpha_\MSbar - \frac{95}{144} \alpha_\MSbar^2 + \frac{1}{16} \alpha_\MSbar^3 - \frac{1}{16} \alpha_\MSbar^4  + \nonumber \\
&& \qquad \qquad \qquad \qquad \qquad \qquad \qquad \left(-\frac{137}{12} + \frac{13}{36} \alpha_\MSbar + \frac{11}{24} \alpha_\MSbar^2 - \frac{3}{8} \alpha_\MSbar^3 \right) \logq + \nonumber \\
&& \qquad \qquad \qquad \qquad \qquad \qquad \qquad \left(\frac{13}{8} + \frac{17}{24} \alpha_\MSbar - \frac{1}{4} \alpha_\MSbar^2 \right) \logqsq + \nonumber \\
&& \qquad \qquad \qquad \qquad \qquad \qquad \qquad \ \left(-3 + 2 \alpha_\MSbar \right) \zeta (3) \Bigg) + \nonumber \\
&& \qquad \qquad \qquad \quad \qquad \qquad \quad \frac{N_f}{N_c} \Bigg(\frac{55}{12} - \logq - 4 \zeta (3) \Bigg) + \nonumber \\
&& \qquad \qquad \qquad \quad \qquad \qquad N_f N_c \Bigg(-\frac{287}{24} + \frac{5}{9} \alpha_\MSbar + \frac{5}{9} \alpha_\MSbar^2 + \nonumber \\
&& \qquad \qquad \qquad \qquad \qquad \qquad \qquad \left(\frac{31}{6} + \frac{2}{9} \alpha_\MSbar - \frac{1}{3} \alpha_\MSbar^2 \right) \logq + \nonumber \\
&& \qquad \qquad \qquad \qquad \qquad \qquad \qquad \left(-\frac{1}{2} - \frac{1}{3} \alpha_\MSbar \right) \logqsq \Bigg) \Bigg] + \mathcal{O} (g_\MSbar^6) \Bigg\}.
\end{eqnarray}

\bigskip

The expressions for the Green's functions with a pair of external quark and antiquark fields and zero-momentum operator insertion ($G^\MSbar_{qi} (q,q)$) are as follows: 

\begin{eqnarray}
&& G^\MSbar_{q1} (q,q) = \qquad \qquad \qquad \qquad \qquad \qquad \qquad \qquad \qquad \qquad \qquad \qquad\qquad \qquad \qquad \qquad \quad \nonumber \\
&& \qquad G_8 \Bigg\{ \frac{g_\MSbar^2}{16 \pi^2} C_F \ \left(\frac{22}{9} - \frac{8}{3} \logq \right) \nonumber \\
&& \qquad \qquad + \frac{g_\MSbar^4}{(16 \pi^2)^2} \Bigg[ - \frac{1177}{216} - \frac{1}{24} \alpha_\MSbar + \left( \frac{181}{36} + \frac{53}{36} \alpha_\MSbar \right) \logq + \nonumber \\
&& \qquad \qquad \qquad \qquad \qquad \ \ \left(- \frac{2}{3} - \frac{4}{3} \alpha_\MSbar \right) \logqsq + \left(- \frac{3}{2} + \frac{1}{2} \alpha_\MSbar \right) \zeta (3) + \nonumber \\
&& \qquad \qquad \qquad \qquad \ \ \frac{1}{N_c^2} \Bigg(-\frac{1615}{324} - \frac{1}{4} \alpha_\MSbar + \left( \frac{10}{3} - \frac{11}{18} \alpha_\MSbar \right) \logq + \nonumber \\
&& \qquad \qquad \qquad \qquad \qquad \ \ \ \ \left(- \frac{8}{9} + \frac{2}{3} \alpha_\MSbar \right) \logqsq + 4 \zeta (3) \Bigg) + \nonumber \\
&& \qquad \qquad \qquad \qquad \ \ \ N_c^2 \Bigg(\frac{6761}{648} + \frac{7}{24} \alpha_\MSbar + \left(- \frac{301}{36} - \frac{31}{36} \alpha_\MSbar \right) \logq + \nonumber \\
&& \qquad \qquad \qquad \qquad \qquad \ \ \ \ \left( \frac{14}{9} + \frac{2}{3} \alpha_\MSbar \right) \logqsq + \left(-\frac{5}{2} - \frac{1}{2} \alpha_\MSbar \right) \zeta (3) \Bigg) + \nonumber \\
&& \qquad \qquad \qquad \quad \ \ N_f C_F \Bigg(- \frac{1022}{81} + \frac{64}{9} \logq - \frac{16}{9} \logqsq  \Bigg) \Bigg] + \mathcal{O} (g_\MSbar^6) \Bigg\} + \nonumber \\
&& \qquad G_9 \Bigg\{ \frac{g_\MSbar^2}{16 \pi^2} C_F \ \left(- \frac{2}{3} \right) \nonumber \\
&& \qquad \qquad + \frac{g_\MSbar^4}{(16 \pi^2)^2} \Bigg[ \frac{71}{24} + \frac{31}{72} \alpha_\MSbar + \left(- \frac{17}{12} - \frac{23}{12} \alpha_\MSbar \right) \logq + \nonumber \\
&& \qquad \qquad \qquad \qquad \qquad \ \ \ \left(- \frac{3}{2} + \frac{1}{2} \alpha_\MSbar \right) \zeta (3) + \nonumber \\
&& \qquad \qquad \qquad \qquad \ \ \frac{1}{N_c^2} \Bigg( \frac{1}{3} - \frac{13}{36} \alpha_\MSbar + \left( \frac{2}{9} + \frac{5}{6} \alpha_\MSbar \right) \logq \Bigg) + \nonumber \\
&& \qquad \qquad \qquad \qquad \ \ N_c^2 \Bigg(- \frac{79}{24} - \frac{5}{72} \alpha_\MSbar + \left( \frac{43}{36} + \frac{13}{12} \alpha_\MSbar \right) \logq + \nonumber \\
&& \qquad \qquad \qquad \qquad \qquad \ \ \ \ \left( \frac{3}{2} - \frac{1}{2} \alpha_\MSbar \right) \zeta (3) \Bigg) + \nonumber \\
&& \qquad \qquad \qquad \quad \ \ N_f C_F \Bigg( \frac{4}{3} - \frac{8}{9} \logq \Bigg) \Bigg] + \mathcal{O} (g_\MSbar^6) \Bigg\},
\end{eqnarray} \\
\vspace*{-1cm}\begin{eqnarray}
&& G^\MSbar_{q2} (q,q) = \qquad \qquad \qquad \qquad \qquad \qquad \qquad \qquad \qquad \qquad \qquad \qquad\qquad \qquad \qquad \qquad \quad \nonumber \\
&& \qquad G_8 \Bigg\{ 1 + \frac{g_\MSbar^2}{16 \pi^2} C_F \ \left(- \frac{31}{9} + \left(\frac{8}{3} - \alpha_\MSbar \right) \logq \right) \nonumber \\
&& \qquad \qquad + \frac{g_\MSbar^4}{(16 \pi^2)^2} \Bigg[ \frac{3593}{432} - \frac{19}{12} \alpha_\MSbar + \frac{3}{16} \alpha_\MSbar^2 + \left(- \frac{305}{72} - \frac{13}{18} \alpha_\MSbar + \frac{1}{8} \alpha_\MSbar^2 \right) \logq + \nonumber \\
&& \qquad \qquad \qquad \qquad \qquad \ \ \ \left(\frac{2}{3} + \frac{23}{24} \alpha_\MSbar - \frac{3}{8} \alpha_\MSbar^2 \right) \logqsq + \alpha_\MSbar \ \zeta (3) + \nonumber \\
&& \qquad \qquad \qquad \qquad \ \ \frac{1}{N_c^2} \Bigg( \frac{8195}{2592} + \frac{3}{8} \alpha_\MSbar - \frac{1}{8} \alpha_\MSbar^2 + \left(- \frac{71}{24} + \frac{31}{36} \alpha_\MSbar \right) \logq + \nonumber \\
&& \qquad \qquad \qquad \qquad \qquad \ \ \ \ \left( \frac{8}{9} - \frac{2}{3} \alpha_\MSbar + \frac{1}{8} \alpha_\MSbar^2 \right) \logqsq - 2 \zeta (3) \Bigg) + \nonumber \\
&& \qquad \qquad \qquad \qquad \ \ N_c^2 \Bigg(- \frac{29753}{2592} + \frac{29}{24} \alpha_\MSbar - \frac{1}{16} \alpha_\MSbar^2 + \nonumber \\
&& \qquad \qquad \qquad \qquad \qquad \ \ \ \ \left( \frac{259}{36} - \frac{5}{36} \alpha_\MSbar - \frac{1}{8} \alpha_\MSbar^2 \right) \logq + \nonumber \\
&& \qquad \qquad \qquad \qquad \qquad \ \ \ \ \left(- \frac{14}{9} - \frac{7}{24} \alpha_\MSbar + \frac{1}{4} \alpha_\MSbar^2 \right) \logqsq + \left( 2 - \alpha_\MSbar \right) \zeta (3) \Bigg) + \nonumber \\
&& \qquad \qquad \qquad \quad \ \ N_f C_F \Bigg( \frac{4097}{324} - \frac{61}{9} \logq + \frac{16}{9} \logqsq  \Bigg) \Bigg] + \mathcal{O} (g_\MSbar^6) \Bigg\} + \nonumber \\
&& \qquad G_9 \Bigg\{ \frac{g_\MSbar^2}{16 \pi^2} C_F \ \left(- \frac{1}{3} - \alpha_\MSbar \right) \nonumber \\
&& \qquad \qquad + \frac{g_\MSbar^4}{(16 \pi^2)^2} \Bigg[ \frac{31}{18} + \frac{13}{9} \alpha_\MSbar + \frac{3}{4} \alpha_\MSbar^2 + \left(- \frac{1}{6} + \frac{5}{12} \alpha_\MSbar - \frac{3}{4} \alpha_\MSbar^2 \right) \logq + \nonumber \\
&& \qquad \qquad \qquad \qquad \qquad \ \ \ \left( \frac{5}{2} - \frac{1}{2} \alpha_\MSbar \right) \zeta (3) + \nonumber \\
&& \qquad \qquad \qquad \qquad \ \ \frac{1}{N_c^2} \Bigg( \frac{37}{12} + \frac{35}{72} \alpha_\MSbar - \frac{1}{8} \alpha_\MSbar^2 + \left(- \frac{2}{9} - \frac{7}{12} \alpha_\MSbar + \frac{1}{4} \alpha_\MSbar^2 \right) \logq - \nonumber \\
&& \qquad \qquad \qquad \qquad \qquad \ \ \ \ 2 \zeta (3) \Bigg) + \nonumber \\
&& \qquad \qquad \qquad \qquad \ \ N_c^2 \Bigg(- \frac{173}{36} - \frac{139}{72} \alpha_\MSbar - \frac{5}{8} \alpha_\MSbar^2 + \left( \frac{7}{18} + \frac{1}{6} \alpha_\MSbar + \frac{1}{2} \alpha_\MSbar^2 \right) \logq + \nonumber \\
&& \qquad \qquad \qquad \qquad \qquad \ \ \ \ \left(- \frac{1}{2} + \frac{1}{2} \alpha_\MSbar \right) \zeta (3) \Bigg) + \nonumber \\
&& \qquad \qquad \qquad \quad \ \ N_f C_F \Bigg( \frac{10}{9} + \frac{2}{9} \logq \Bigg) \Bigg] + \mathcal{O} (g_\MSbar^6) \Bigg\},
\end{eqnarray} \\
\vspace*{-1cm}\begin{eqnarray}
&& G^\MSbar_{q3} (q,q) = G^\MSbar_{q4} (q,q) = \qquad \qquad \qquad \qquad \qquad \qquad \qquad \qquad \qquad \qquad \qquad \qquad\qquad \quad \ \nonumber \\
&& \qquad G_8 \Bigg\{ \frac{g_\MSbar^2}{16 \pi^2} C_F \ \left( 1 + \alpha_\MSbar \right) \nonumber \\
&& \qquad \qquad + \frac{g_\MSbar^4}{(16 \pi^2)^2} \Bigg[- \frac{553}{72} - \frac{13}{8} \alpha_\MSbar - \frac{3}{4} \alpha_\MSbar^2 + \nonumber \\
&& \qquad \qquad \qquad \qquad \qquad \ \ \ \ \left( \frac{19}{12} + \alpha_\MSbar + \frac{3}{4} \alpha_\MSbar^2 \right) \logq + 3 \zeta (3) + \nonumber \\
&& \qquad \qquad \qquad \qquad \ \ \frac{1}{N_c^2} \Bigg( \frac{5}{3} - \frac{1}{8} \alpha_\MSbar + \frac{1}{8} \alpha_\MSbar^2 + \nonumber \\
&& \qquad \qquad \qquad \qquad \qquad \ \ \ \ \left(- \frac{1}{4} \alpha_\MSbar - \frac{1}{4} \alpha_\MSbar^2 \right) \logq - 2 \zeta (3) \Bigg) + \nonumber \\
&& \qquad \qquad \qquad \qquad \ \ \ N_c^2 \Bigg( \frac{433}{72} + \frac{7}{4} \alpha_\MSbar + \frac{5}{8} \alpha_\MSbar^2 + \nonumber \\
&& \qquad \qquad \qquad \qquad \qquad \ \ \ \ \left(- \frac{19}{12} - \frac{3}{4} \alpha_\MSbar - \frac{1}{2} \alpha_\MSbar^2 \right) \logq - \zeta (3) \Bigg) + \nonumber \\
&& \qquad \qquad \qquad \quad \ \ \ N_f C_F \Bigg(-\frac{16}{9} + \frac{2}{3} \logq \Bigg) \Bigg] + \mathcal{O} (g_\MSbar^6) \Bigg\} + \nonumber \\
&& \qquad G_9 \Bigg\{ \frac{g_\MSbar^2}{16 \pi^2} C_F \ \left(1 - \alpha_\MSbar \right) \nonumber \\
&& \qquad \qquad + \frac{g_\MSbar^4}{(16 \pi^2)^2} \Bigg[ \frac{5}{72} + \frac{13}{8} \alpha_\MSbar + \alpha_\MSbar^2 + \left( \frac{19}{12} - \frac{3}{4} \alpha_\MSbar^2 \right) \logq - \zeta (3) + \nonumber \\
&& \qquad \qquad \qquad \qquad \ \ \frac{1}{N_c^2} \Bigg(- \frac{8}{3} - \frac{1}{8} \alpha_\MSbar - \frac{3}{8} \alpha_\MSbar^2 + \nonumber \\
&& \qquad \qquad \qquad \qquad \qquad \ \ \ \ \left(- \frac{1}{4} \alpha_\MSbar + \frac{1}{4} \alpha_\MSbar^2 \right) \logq + 2 \zeta (3) \Bigg) + \nonumber \\
&& \qquad \qquad \qquad \qquad \ \ \ N_c^2 \Bigg( \frac{187}{72} - \frac{3}{2} \alpha_\MSbar - \frac{5}{8} \alpha_\MSbar^2 + \nonumber \\
&& \qquad \qquad \qquad \qquad \qquad \ \ \ \ \left(- \frac{19}{12} + \frac{1}{4} \alpha_\MSbar + \frac{1}{2} \alpha_\MSbar^2 \right) \logq - \zeta (3) \Bigg) + \nonumber \\
&& \qquad \qquad \qquad \quad \ \ \ N_f C_F \Bigg(- \frac{4}{9} + \frac{2}{3} \logq \Bigg) \Bigg] + \mathcal{O} (g_\MSbar^6) \Bigg\}, 
\end{eqnarray}
\begin{eqnarray}
&& G^\MSbar_{q5} (q,q) = 0. \qquad \qquad \qquad \qquad \qquad \qquad \qquad \qquad \qquad \qquad \qquad \qquad\qquad \qquad \qquad \qquad
\label{MSbar_last}
\end{eqnarray}

\subsection{Renormalization factors in the $\MSbar$ scheme}

Here, we provide our results for the renormalization factors of operators $O_i$, $(i=1, 2, \ldots, 5)$ in (DR, $\MSbar$), as a Laurent series in $\varepsilon \equiv (4-d)/2$, 
\begin{equation}
Z^{\overline{\rm MS},{\rm DR}}_{ij} = \delta_{ij} + {[z^{\overline{\rm MS},{\rm DR}}_{1,-1}]}_{ij} \frac{g_{\overline{\rm MS}}^2}{16 \pi^2 \varepsilon} + {[z^{\overline{\rm MS},{\rm DR}}_{2,-2}]}_{ij} \frac{g_{\overline{\rm MS}}^4}{{(16 \pi^2)}^2 \varepsilon^2} + {[z^{\overline{\rm MS},{\rm DR}}_{2,-1}]}_{ij} \frac{g_{\overline{\rm MS}}^4}{{(16 \pi^2)}^2 \varepsilon} + \mathcal{O} (g_{\overline{\rm MS}}^6),
\end{equation}
where 
\begin{equation}
z^{\overline{\rm MS},{\rm DR}}_{1,-1} = \begin{pmatrix}
\phantom{-}a_1 & \phantom{-}b_1 & \phantom{-}0 & \phantom{-}c_1 & -c_1 \\
-a_1 & -b_1 & \phantom{-}0 & \phantom{-}0 & \phantom{-}0 \\
\phantom{-}0 & \phantom{-}0 & \phantom{-}0 & -c_1 & \phantom{-}c_1 \\
\phantom{-}0 & \phantom{-}0 & \phantom{-}0 & -c_1 & \phantom{-}c_1 \\
\phantom{-}0 & \phantom{-}0 & \phantom{-}0 & \phantom{-}0 & \phantom{-}0
\end{pmatrix}, 
\end{equation} 
\begin{equation}
z^{\overline{\rm MS},{\rm DR}}_{2,-2} = \begin{pmatrix}
\phantom{-}a_2 & \phantom{-}b_2 & \phantom{-}0 & \phantom{-}c_2 & -c_2 \\
-a_2 & -b_2 & \phantom{-}0 & \phantom{-}d_2 & -d_2 \\
\phantom{-}0 & \phantom{-}0 & \phantom{-}0 & \phantom{-}e_2 & -e_2 \\
\phantom{-}0 & \phantom{-}0 & \phantom{-}0 & \phantom{-}e_2 & -e_2 \\
\phantom{-}0 & \phantom{-}0 & \phantom{-}0 & \phantom{-}0 & \phantom{-}0
\end{pmatrix}, \quad 
z^{\overline{\rm MS},{\rm DR}}_{2,-1} = \begin{pmatrix}
\phantom{-}a_3 & \phantom{-}b_3 & \phantom{-}0 & \phantom{-}c_3 & -c_3 \\
-a_3 & -b_3 & \phantom{-}0 & \phantom{-}d_3 & -d_3 \\
\phantom{-}0 & \phantom{-}0 & \phantom{-}0 & \phantom{-}e_3 & -e_3 \\
\phantom{-}0 & \phantom{-}0 & \phantom{-}0 & \phantom{-}e_3 & -e_3 \\
\phantom{-}0 & \phantom{-}0 & \phantom{-}0 & \phantom{-}0 & \phantom{-}0
\end{pmatrix},
\end{equation} 
and
\begin{equation}
a_1 = \frac{2 N_f}{3}, \qquad b_1 = - \frac{8 C_F}{3}, \qquad c_1 = - \frac{N_c}{2},
\end{equation}
\begin{equation}
a_2 = -\frac{N_f}{9} \left( \frac{4}{N_c} + 7 N_c - 4 N_f \right), \qquad b_2 = \frac{4 C_F}{9} \left( \frac{4}{N_c} + 7 N_c - 4 N_f \right), 
\end{equation}
\begin{equation}
c_2 = \frac{N_c}{24} \left(19 N_c - 8 N_f \right), \qquad d_2 = \frac{N_f N_c}{6}, \qquad e_2 = - \frac{N_c}{24} \left(19 N_c - 4 N_f \right),
\end{equation}
\begin{equation}
a_3 = -\frac{N_f}{54} \left( \frac{37}{N_c} - 72 N_c \right), \qquad b_3 = -\frac{4 C_F}{27} \left( \frac{7}{N_c} + 40 N_c - 13 N_f \right), 
\end{equation}
\begin{equation}
c_3 = -\frac{N_c}{144} \left[15 \left(6 + \alpha_\MSbar \right) N_c - 56 N_f \right], \qquad d_3 = -2 \frac{N_f N_c}{9}, 
\end{equation}
\begin{equation}
e_3 = \frac{N_c}{48} \left[5 \left(6 + \alpha_\MSbar \right) N_c - 8 N_f \right].
\end{equation}
As was expected, the mixing matrix is block triangular. Also, there is no mixing between $O_i$ ($i \neq 3$) and $O_3$; the third column has zero elements except from the diagonal element ($i=3$) which equals one. Furthermore, operators $O_1$, $O_2$, $O_3$ mix with the linear combination $O_4 - O_5$ and not with $O_4$ and $O_5$ separately. This becomes apparent when one replaces, e.g., $O_4$ with $O_4 - O_5$ in constructing the mixing matrix [see Eq. \eqref{Z_O}]. Moreover, Eqs. (\ref{cons1} -- \ref{triag}) are automatically fulfilled. In conclusion, our results agree with the theoretical analysis given in Sec. \ref{TheoreticalAnalysis}.

\subsection{Conversion factors}
\label{Conversion factors}

Here, we present our results for the conversion factors $C_{ij}^{\MSbar, {\rm RI}'}$ between the different versions of RI$'$ and $\MSbar$ scheme. For the sake of brevity, we provide only our resulting expressions for the RI$'_2$ scheme and its ``conserved'' version, RI${'_2}^{\rm cons}$, while the conversion factors for RI$'_1$ can be extracted from Eqs. \eqref{MSbar_first} -- \eqref{MSbar_last}. Our results depend on two renormalization scales: the RI$'$ scale $\bar{q}$ and the $\MSbar$ scale $\bar{\mu}$; we have chosen to keep these two scales distinct, for wider applicability. Note that, whereas the matrix $Z^{\MSbar, {\rm DR}}$ is necessarily block triangular, no such condition applies to the matrix $C^{\MSbar, {\rm RI}'}$.

\bigskip

The expressions for the conversion factors $C_{ij}^{\MSbar, {\rm RI}'_2}$ $(i,j = 1, 2, 6)$ between RI$'_2$ and $\MSbar$ are as follows:
\begin{eqnarray}
&& C_{11}^{\MSbar,{\rm RI}'_2} = \qquad \qquad \qquad \qquad \qquad \qquad \qquad \qquad \qquad \qquad \qquad \qquad\qquad \qquad \qquad \qquad \quad \nonumber \\
&& \qquad 1 + \frac{g_\MSbar^2}{16 \pi^2} \Bigg[ N_c \Bigg( \frac{5}{12} - \frac{3}{2} \alpha_\MSbar - \frac{1}{4} \alpha_\MSbar^2 \Bigg) + N_f \Bigg(- \frac{10}{9} + \frac{2}{3} \logqb \Bigg) \Bigg] \nonumber \\
&& \qquad \ \ + \frac{g_\MSbar^4}{{(16 \pi^2)}^2 } \Bigg[ N_c^2 \Bigg(- \frac{14483}{2160} - \frac{1697}{240} \alpha_\MSbar + \frac{139}{360} \alpha_\MSbar^2 + \frac{3}{16} \alpha_\MSbar^3 + \frac{1}{16} \alpha_\MSbar^4 + \nonumber \\
&& \qquad \qquad \qquad \qquad \qquad \left( - \frac{173}{72} + \frac{7}{4} \alpha_\MSbar + \frac{11}{24} \alpha_\MSbar^2 + \frac{1}{4} \alpha_\MSbar^3 \right) \logqb + \nonumber \\
&& \qquad \qquad \qquad \qquad \qquad \left( \frac{69}{10} + \frac{7}{5} \alpha_\MSbar - \frac{1}{5} \alpha_\MSbar^2 \right) \zeta (3) \Bigg) + \nonumber \\
&& \qquad \qquad \qquad \quad \ \frac{N_f}{N_c} \Bigg( \frac{587}{162} - \frac{7}{9} \logqb - \frac{4}{9} \logqbsq - 4 \zeta (3) \Bigg) + \nonumber \\
&& \qquad \qquad \qquad \ N_f N_c \Bigg(- \frac{1019}{162} + \frac{5}{3} \alpha_\MSbar + \left( \frac{181}{27} - \alpha_\MSbar \right) \logqb \nonumber \\
&& \qquad \qquad \qquad \qquad \qquad - \frac{7}{9} \logqbsq - 6 \zeta (3) \Bigg) + \nonumber \\
&& \qquad \qquad \qquad \quad \ N_f^2 \Bigg( \frac{100}{81} - \frac{40}{27} \logqb + \frac{4}{9} \logqbsq \Bigg) \Bigg] + \mathcal{O} (g_\MSbar^6),
\end{eqnarray}
\begin{eqnarray}
&& C_{12}^{\MSbar,{\rm RI}'_2} = \qquad \qquad \qquad \qquad \qquad \qquad \qquad \qquad \qquad \qquad \qquad \qquad\qquad \qquad \qquad \qquad \quad \nonumber \\
&& \qquad \frac{g_\MSbar^2}{16 \pi^2} C_F \Bigg( \frac{16}{9} - \frac{8}{3} \logqb \Bigg) \nonumber \\
&& \qquad + \frac{g_\MSbar^4}{{(16 \pi^2)}^2 } C_F \Bigg[ \frac{1}{N_c} \Bigg(\frac{1507}{162} + \frac{19}{9} \alpha_\MSbar - \frac{64}{9} \logqb + \nonumber \\
&& \qquad \qquad \qquad \qquad \quad \ \ \ \ \frac{16}{9} \logqbsq - 8 \zeta (3) \Bigg) + \nonumber \\
&& \qquad \qquad \qquad \quad \ \ \ \ N_c \Bigg( \frac{1157}{81} - \frac{4}{9} \alpha_\MSbar - \frac{43}{3} \logqb + \frac{28}{9} \logqbsq + \nonumber \\
&& \qquad \qquad \qquad \quad \qquad \ \ \ \ \left(-2 - 2 \alpha_\MSbar \right) \zeta (3) \Bigg) + \nonumber \\
&& \qquad \qquad \qquad \quad \ \ \ \ C_F \ \frac{16}{3} \alpha_\MSbar \logqb + \nonumber \\
&& \qquad \qquad \qquad \quad \ \ \ \ N_f \Bigg(- \frac{914}{81} + \frac{56}{9} \logqb - \frac{16}{9} \logqbsq \Bigg) \Bigg] + \mathcal{O} (g_\MSbar^6),
\end{eqnarray}
\begin{eqnarray}
&& C_{16}^{\MSbar,{\rm RI}'_2} = \qquad \qquad \qquad \qquad \qquad \qquad \qquad \qquad \qquad \qquad \qquad \qquad\qquad \qquad \qquad \qquad \quad \nonumber \\
&& \qquad \frac{g_\MSbar^2}{16 \pi^2} N_c \Bigg( \frac{5}{6} - \frac{1}{2} \alpha_\MSbar - \frac{1}{2} \logqb \Bigg) \nonumber \\
&& \qquad + \frac{g_\MSbar^4}{{(16 \pi^2)}^2 } \Bigg[ N_c^2 \Bigg( \frac{3469}{1080} - \frac{991}{480} \alpha_\MSbar + \frac{11}{60} \alpha_\MSbar^2 + \frac{1}{8} \alpha_\MSbar^3 + \nonumber \\
&& \qquad \qquad \qquad \qquad \quad \left( - \frac{319}{72} + \frac{5}{12} \alpha_\MSbar + \frac{1}{4} \alpha_\MSbar^2 \right) \logqb + \frac{19}{24} \logqbsq + \nonumber \\
&& \qquad \qquad \qquad \qquad \quad \left( \frac{47}{20} + \frac{3}{5} \alpha_\MSbar - \frac{1}{20} \alpha_\MSbar^2 \right) \zeta (3) \Bigg) + \nonumber \\
&& \qquad \qquad \qquad \ \ \frac{N_f}{N_c} \Bigg(- \frac{2}{9} - \frac{4}{9} \logqb \Bigg) + \nonumber \\
&& \qquad \qquad \quad \ \ N_f N_c \Bigg(- \frac{257}{108} + \frac{5}{9} \alpha_\MSbar + \left( \frac{7}{3} - \frac{1}{3} \alpha_\MSbar \right) \logqb \nonumber \\
&& \qquad \qquad \qquad \qquad \qquad - \frac{1}{3} \logqbsq - 2 \zeta (3) \Bigg) \Bigg] + \mathcal{O} (g_\MSbar^6),
\end{eqnarray}
\begin{eqnarray}
&& C_{21}^{\MSbar,{\rm RI}'_2} = \qquad \qquad \qquad \qquad \qquad \qquad \qquad \qquad \qquad \qquad \qquad \qquad\qquad \qquad \qquad \qquad \quad \nonumber \\
&& \qquad \frac{g_\MSbar^2}{16 \pi^2} N_f \Bigg( \frac{4}{9} - \frac{2}{3} \logqb \Bigg) \nonumber \\
&& \qquad + \frac{g_\MSbar^4}{{(16 \pi^2)}^2 } N_f \Bigg[ \frac{1}{N_c} \Bigg(- \frac{425}{162} + \frac{7}{9}  \logqb + \frac{4}{9} \logqbsq + 4 \zeta (3) \Bigg) + \nonumber \\
&& \qquad \qquad \qquad \qquad \ N_c \Bigg( \frac{3551}{1620} - \frac{7}{12} \alpha_\MSbar - \frac{1}{9} \alpha_\MSbar^2 + \nonumber \\
&& \qquad \qquad \qquad \qquad \qquad \ \left(- \frac{215}{54} + \alpha_\MSbar + \frac{1}{6} \alpha_\MSbar^2 \right) \logqb + \nonumber \\
&& \qquad \qquad \qquad \qquad \qquad \ \ \frac{7}{9} \logqbsq + \frac{28}{5} \zeta (3) \Bigg) + \nonumber \\
&& \qquad \qquad \qquad \qquad \ N_f \Bigg( -\frac{40}{81} + \frac{28}{27} \logqb - \frac{4}{9} \logqbsq \Bigg) \Bigg] + \mathcal{O} (g_\MSbar^6),
\end{eqnarray}
\begin{eqnarray}
&& C_{22}^{\MSbar,{\rm RI}'_2} = \qquad \qquad \qquad \qquad \qquad \qquad \qquad \qquad \qquad \qquad \qquad \qquad\qquad \qquad \qquad \qquad \quad \nonumber \\
&& \qquad 1 + \frac{g_\MSbar^2}{16 \pi^2} C_F \Bigg(- \frac{34}{9} - 2 \alpha_\MSbar + \frac{8}{3} \logqb \Bigg) \nonumber \\
&& \qquad \ \ + \frac{g_\MSbar^4}{{(16 \pi^2)}^2 } C_F \Bigg[ \frac{1}{N_c} \Bigg(- \frac{1037}{81} - \frac{65}{18} \alpha_\MSbar - \frac{1}{2} \alpha_\MSbar^2 + \nonumber \\
&& \qquad \qquad \qquad \qquad \quad \ \ \ \ \ \ \left( \frac{64}{9} + \frac{8}{3} \alpha_\MSbar \right) \logqb - \frac{16}{9} \logqbsq + 8 \zeta (3) \Bigg) + \nonumber \\
&& \qquad \qquad \qquad \quad \ \ \ \ \ \ N_c \Bigg(- \frac{3443}{81} - \frac{109}{18} \alpha_\MSbar - \frac{3}{2} \alpha_\MSbar^2 + \nonumber \\
&& \qquad \qquad \qquad \quad \qquad \ \ \ \ \ \ \ \left( \frac{62}{3} + \frac{1}{3} \alpha_\MSbar + \alpha_\MSbar^2 \right) \logqb - \frac{28}{9} \logqbsq + \nonumber \\
&& \qquad \qquad \qquad \quad \qquad \ \ \ \ \ \ \ \ \left(6 + 2 \alpha_\MSbar \right) \zeta (3) \Bigg) + \nonumber \\
&& \qquad \qquad \qquad \quad \ \ \ \ \ \ N_f \Bigg( \frac{1256}{81} - \frac{68}{9} \logqb + \frac{16}{9} \logqbsq \Bigg) \Bigg] + \mathcal{O} (g_\MSbar^6),
\end{eqnarray}
\begin{eqnarray}
&& C_{26}^{\MSbar,{\rm RI}'_2} = \qquad \qquad \qquad \qquad \qquad \qquad \qquad \qquad \qquad \qquad \qquad \qquad\qquad \qquad \qquad \qquad \quad \nonumber \\
&& \qquad \frac{g_\MSbar^2}{16 \pi^2} N_f \Bigg(- \frac{1}{3} \Bigg) \nonumber \\
&& \qquad + \frac{g_\MSbar^4}{{(16 \pi^2)}^2 } N_f \Bigg[ \frac{1}{N_c} \Bigg( \frac{13}{18} + \frac{4}{9}  \logqb \Bigg) + \nonumber \\
&& \qquad \qquad \qquad \qquad \ N_c \Bigg(- \frac{157}{90} - \frac{1}{18} \alpha_\MSbar + \frac{1}{12} \alpha_\MSbar^2 + \left(- \frac{2}{9} + \frac{1}{3} \alpha_\MSbar \right) \logqb + \nonumber \\
&& \qquad \qquad \qquad \qquad \qquad \ \ \frac{1}{6} \logqbsq + \frac{12}{5} \zeta (3) \Bigg) + \nonumber \\
&& \qquad \qquad \qquad \qquad \ N_f \Bigg( \frac{10}{27} - \frac{2}{9} \logqb \Bigg) \Bigg] + \mathcal{O} (g_\MSbar^6),
\end{eqnarray}
\begin{eqnarray}
&& C_{61}^{\MSbar,{\rm RI}'_2} = \qquad \qquad \qquad \qquad \qquad \qquad \qquad \qquad \qquad \qquad \qquad \qquad\qquad \qquad \qquad \qquad \quad \nonumber \\
&& \qquad \frac{g_\MSbar^2}{16 \pi^2} N_c \Bigg( \frac{7}{4} + \alpha_\MSbar + \frac{1}{4} \alpha_\MSbar^2 \Bigg) \nonumber \\
&& \qquad + \frac{g_\MSbar^4}{{(16 \pi^2)}^2 } N_c \Bigg[ N_c \Bigg( \frac{17251}{720} + \frac{1547}{240}  \alpha_\MSbar - \frac{199}{360} \alpha_\MSbar^2 + \frac{1}{16} \alpha_\MSbar^3 - \frac{1}{16} \alpha_\MSbar^4 + \nonumber \\
&& \qquad \qquad \qquad \qquad \qquad \ \left(-\frac{133}{24} - \alpha_\MSbar - \frac{5}{24} \alpha_\MSbar^2 - \frac{1}{4} \alpha_\MSbar^3 \right) \logqb + \nonumber \\
&& \qquad \qquad \qquad \qquad \qquad \ \ \left(- \frac{69}{10} - \frac{7}{5} \alpha_\MSbar + \frac{1}{5} \alpha_\MSbar^2 \right) \zeta (3) \Bigg) + \nonumber \\
&& \qquad \qquad \qquad \quad \ \ \ N_f \Bigg(-\frac{949}{180} - \frac{13}{12} \alpha_\MSbar + \frac{5}{18} \alpha_\MSbar^2 + \left( \frac{7}{6} - \frac{1}{6} \alpha_\MSbar^2 \right) \logqb + \nonumber \\
&& \qquad \qquad \qquad \qquad \qquad \ \ \frac{2}{5} \zeta (3) \Bigg) \Bigg] + \mathcal{O} (g_\MSbar^6),
\end{eqnarray}
\begin{eqnarray}
&& C_{62}^{\MSbar,{\rm RI}'_2} = \qquad \qquad \qquad \qquad \qquad \qquad \qquad \qquad \qquad \qquad \qquad \qquad\qquad \qquad \qquad \qquad \quad \nonumber \\
&& \qquad 1 + \frac{g_\MSbar^2}{16 \pi^2} C_F \ 2 \nonumber \\
&& \qquad \ \ + \frac{g_\MSbar^4}{{(16 \pi^2)}^2 } C_F \Bigg[ \frac{1}{N_c} \Bigg( 2 + \frac{3}{2} \alpha_\MSbar + \frac{1}{2} \alpha_\MSbar^2 \Bigg) + \nonumber \\
&& \qquad \qquad \qquad \quad \ \ \ \ \ \ N_c \Bigg(\frac{155}{9} - \frac{1}{2} \alpha_\MSbar - \frac{19}{3} \logqb - 4 \zeta (3) \Bigg) + \nonumber \\
&& \qquad \qquad \qquad \quad \ \ \ \ \ \ N_f \Bigg(- \frac{20}{9} + \frac{4}{3} \logqb \Bigg) \Bigg] + \mathcal{O} (g_\MSbar^6),
\end{eqnarray}
\begin{eqnarray}
&& C_{66}^{\MSbar,{\rm RI}'_2} = \qquad \qquad \qquad \qquad \qquad \qquad \qquad \qquad \qquad \qquad \qquad \qquad\qquad \qquad \qquad \qquad \quad \nonumber \\
&& \qquad 1 + \frac{g_\MSbar^2}{16 \pi^2} N_c \Bigg( \frac{1}{4} + \frac{1}{4} \alpha_\MSbar + \frac{1}{2} \logqb \Bigg) \nonumber \\
&& \qquad + \frac{g_\MSbar^4}{{(16 \pi^2)}^2 } N_c \Bigg[ N_c \Bigg( \frac{3899}{720} + \frac{841}{480}  \alpha_\MSbar - \frac{4}{15} \alpha_\MSbar^2 + \nonumber \\
&& \qquad \qquad \qquad \qquad \qquad \ \left( \frac{11}{24} - \frac{1}{24} \alpha_\MSbar - \frac{1}{8} \alpha_\MSbar^2 \right) \logqb - \nonumber \\
&& \qquad \qquad \qquad \qquad \qquad \ \ \frac{19}{24} \logqbsq + \left(- \frac{47}{20} - \frac{3}{5} \alpha_\MSbar + \frac{1}{20} \alpha_\MSbar^2 \right) \zeta (3) \Bigg) + \nonumber \\
&& \qquad \qquad \qquad \quad \ \ \ N_f \Bigg(- \frac{101}{180} - \frac{1}{2} \alpha_\MSbar - \frac{1}{6} \logqb + \nonumber \\
&& \qquad \qquad \qquad \qquad \qquad \ \ \frac{1}{6} \logqbsq - \frac{2}{5} \zeta (3) \Bigg) \Bigg] + \mathcal{O} (g_\MSbar^6).
\end{eqnarray}

\bigskip

The expression for the conversion factors $C_{ij}^{\MSbar,{\rm RI}{'_2}^{\rm cons}}$ $(i,j = 1, 2, 6)$ between RI${'}_2^{\ {\rm cons}}$ and $\MSbar$ is as follows: 
\begin{equation}
 C_{ij}^{\MSbar,{\rm RI}{'_2}^{\rm cons}} = \ C_{ij}^{\MSbar,{\rm RI}'_2} + \delta C_{ij},
\end{equation}
where
\begin{eqnarray}
\delta C_{11} &=& \frac{g_\MSbar^4}{{(16 \pi^2)}^2 } N_c \Bigg[ N_c \Bigg(- \frac{65}{36} + \frac{3}{2} \alpha_\MSbar - \frac{1}{4} \alpha_\MSbar^2 + \left(\frac{13}{12} - \frac{1}{4} \alpha_\MSbar \right) \logqb \Bigg) + \nonumber \\
&& \qquad \qquad \quad N_f \Bigg( \frac{5}{9} - \frac{1}{3} \alpha_\MSbar - \frac{1}{3} \logqb \Bigg) \Bigg] + \mathcal{O} (g_\MSbar^6), \\
\delta C_{12} &=& \frac{g_\MSbar^4}{{(16 \pi^2)}^2 } C_F \Bigg[ \frac{1}{N_c} \Bigg(- \frac{8}{3} \alpha_\MSbar \logqb \Bigg) + \nonumber \\
&& \qquad \qquad \quad \ \ N_c \Bigg( \frac{5}{3} \alpha_\MSbar - \alpha_\MSbar^2 + \frac{5}{3} \alpha_\MSbar \logqb \Bigg) + \nonumber \\
&& \qquad \qquad \quad \ \ C_F \Bigg(- \frac{16}{3} \alpha_\MSbar \logqb \Bigg) \Bigg] + \mathcal{O} (g_\MSbar^6), \\
\delta C_{21} &=& \frac{g_\MSbar^4}{{(16 \pi^2)}^2 } N_f \Bigg[ N_c \Bigg( \frac{13}{18} - \frac{1}{6} \alpha_\MSbar \Bigg) + N_f \Bigg(- \frac{2}{9} \Bigg) \Bigg] + \mathcal{O} (g_\MSbar^6), \\
\delta C_{22} &=& \frac{g_\MSbar^4}{{(16 \pi^2)}^2 } C_F N_f \Bigg(- \frac{2}{3} \alpha_\MSbar \Bigg) + \mathcal{O} (g_\MSbar^6),
\end{eqnarray}
\begin{eqnarray}
\delta C_{61} &=& \frac{g_\MSbar^2}{16 \pi^2} \Bigg[ N_c \Bigg(- \frac{13}{6} + \frac{1}{2} \alpha_\MSbar \Bigg) + N_f \frac{2}{3} \Bigg] \nonumber \\
&& + \frac{g_\MSbar^4}{{(16 \pi^2)}^2 } \Bigg[ N_c^2 \Bigg(- \frac{3337}{216} - \frac{7}{8} \alpha_\MSbar + \frac{5}{12} \alpha_\MSbar^2 - \frac{1}{4} \alpha_\MSbar^3 + \nonumber \\
&& \qquad \qquad \qquad \ \ \ \left(\frac{247}{36} - \frac{1}{2} \alpha_\MSbar - \frac{1}{4} \alpha_\MSbar^2 \right) \logqb \Bigg) + \nonumber \\
&& \qquad \qquad \ \ \frac{N_f}{N_c} \ (-1) + \nonumber \\
&& \qquad \quad \ \ N_f N_c \Bigg( \frac{437}{54} + \frac{1}{2} \alpha_\MSbar - \frac{1}{6} \alpha_\MSbar^2 - \frac{32}{9} \logqb \Bigg) + \nonumber \\
&& \qquad \qquad \ \ N_f^2 \Bigg(- \frac{14}{27} + \frac{4}{9} \logqb \Bigg) \Bigg] + \mathcal{O} (g_\MSbar^6), \\
\delta C_{62} &=& \frac{g_\MSbar^2}{16 \pi^2} C_F 2 \alpha_\MSbar \nonumber \\
&& + \frac{g_\MSbar^4}{{(16 \pi^2)}^2 } C_F \Bigg[ \frac{1}{N_c} \Bigg( \frac{3}{2} \Bigg) + \nonumber \\
&& \qquad \quad \ \ \ \qquad \ N_c \Bigg( 11 + \frac{16}{3} \alpha_\MSbar + \frac{5}{2} \alpha_\MSbar^2 + \left(- 2 \alpha_\MSbar - \alpha_\MSbar^2 \right) \logqb \Bigg) + \nonumber \\
&& \qquad \quad \ \ \ \qquad \ N_f \Bigg(- 2 + \frac{2}{3} \alpha_\MSbar \Bigg) \Bigg] + \mathcal{O} (g_\MSbar^6), \\
\delta C_{i6} &=& \frac{1}{2} \delta C_{i1}, \quad (i=1, 2, 6).
\end{eqnarray} 

\bigskip

For completeness, we also provide the conversion factors of the gluon and fermion fields, in terms of arbitrary RI$'$ and $\MSbar$ scales. These factors are in agreement with the well-known (in the literature) results for the case $\bar{q} = \bar{\mu}$ (see, e.g., \cite{Gracey:2003yr}).
\vspace{-1cm}
\begin{eqnarray}
C_A^{\MSbar,{\rm RI}'} &=& 1 + \frac{g_\MSbar^2}{16 \pi^2} \Bigg[ N_c \Bigg(- \frac{97}{36} - \frac{1}{2} \alpha_\MSbar - \frac{1}{4} \alpha_\MSbar^2 + \left( \frac{13}{6} - \frac{1}{2} \alpha_\MSbar \right) \logqb \Bigg) + \nonumber \\
&& \qquad \qquad \ \ N_f \Bigg( \frac{10}{9} - \frac{2}{3} \logqb \Bigg) \Bigg] \nonumber \\
&& \ \ + \frac{g_\MSbar^4}{{(16 \pi^2)}^2 } \Bigg[ N_c^2 \Bigg(- \frac{2381}{96} + \frac{463}{288} \alpha_\MSbar + \frac{95}{144} \alpha_\MSbar^2 - \frac{1}{16} \alpha_\MSbar^3 + \frac{1}{16} \alpha_\MSbar^4 + \nonumber \\
&& \qquad \qquad \qquad \quad \ \ \left( \frac{137}{12} - \frac{13}{36} \alpha_\MSbar - \frac{11}{24} \alpha_\MSbar^2 + \frac{3}{8} \alpha_\MSbar^3 \right) \logqb + \nonumber \\
&& \qquad \qquad \qquad \quad \ \ \left(- \frac{13}{8} - \frac{17}{24} \alpha_\MSbar + \frac{1}{4} \alpha_\MSbar^2 \right) \logqbsq + \nonumber \\
&& \qquad \qquad \qquad \quad \ \ \ \left( 3 - 2 \alpha_\MSbar \right) \zeta (3) \Bigg) + \nonumber \\
&& \qquad \qquad \quad \ \frac{N_f}{N_c} \Bigg(- \frac{55}{12} + \logqb + 4 \zeta (3) \Bigg) + \nonumber \\
&& \qquad \qquad \ N_f N_c \Bigg( \frac{287}{24} - \frac{5}{9} \alpha_\MSbar - \frac{5}{9} \alpha_\MSbar^2 + \nonumber \\
&& \qquad \qquad \qquad \quad \ \ \left(- \frac{31}{6} - \frac{2}{9} \alpha_\MSbar + \frac{1}{3} \alpha_\MSbar^2 \right) \logqb + \nonumber \\
&& \qquad \qquad \qquad \quad \ \ \left( \frac{1}{2} + \frac{1}{3} \alpha_\MSbar \right) \logqbsq \Bigg) \Bigg] + \mathcal{O} (g_\MSbar^6),
\end{eqnarray}
\vspace{-0.2cm}
\begin{eqnarray}
 C_\psi^{\MSbar,{\rm RI}'} &=& 1 + \frac{g_\MSbar^2}{16 \pi^2} C_F \Bigg( \alpha_\MSbar - \alpha_\MSbar \logqb \Bigg) + \nonumber \\
&& \ \ + \frac{g_\MSbar^4}{{(16 \pi^2)}^2 } C_F \Bigg[ N_c \Bigg( \frac{41}{4} + \frac{13}{2} \alpha_\MSbar + \frac{9}{8} \alpha_\MSbar^2 + \nonumber \\
&& \qquad \qquad \qquad \quad \ \ \ \ \ \left(- \frac{25}{4} - \frac{7}{2} \alpha_\MSbar - \frac{3}{4} \alpha_\MSbar^2 \right) \logqb + \nonumber \\
&& \qquad \qquad \qquad \quad \ \ \ \ \ \left( \frac{3}{4} \alpha_\MSbar + \frac{1}{4} \alpha_\MSbar^2 \right) \logqbsq + \nonumber \\
&& \qquad \qquad \qquad \quad \ \ \ \ \ \ \ \left(- 3 - 3 \alpha_\MSbar \right) \zeta (3) \Bigg) + \nonumber \\
&& \qquad \qquad \quad \ \ \ \ \ N_f \Bigg(- \frac{7}{4} + \logqb \Bigg) + \nonumber \\
&& \qquad \qquad \quad \ \ \ \ \ C_F \Bigg(- \frac{5}{8} + \left( \frac{3}{2} - \alpha_\MSbar^2 \right) \logqb + \frac{1}{2} \alpha_\MSbar^2 \logqbsq \Bigg) \Bigg] \nonumber \\
&& \ \ + \mathcal{O} (g_\MSbar^6).
\end{eqnarray}

\section{Nonperturbative renormalization}
\label{Nonperturbative renormalization}

The construction of a complete nonperturbative renormalization program, which can eliminate operator-mixing effects, is a difficult task; some well-known complications involve power-divergent mixing of lower-dimensional operators, as well as additional, finite mixing contributions associated with the reduction of rotational to hypercubic invariance.

\bigskip

Additional complications arise when gauge-variant operators (BRST variations and EOM operators) are included in the set of operators which mix. Such operators, typically, contain ghost fields and/or gauge-fixing terms, which are defined in perturbation theory, and their study is not obvious in a nonperturbative context. 

\bigskip

There are various approaches, used in the literature, for the study of operator mixing on the lattice. The first one is the perturbative approach, where the renormalization factors are extracted by lattice perturbation theory (see, e.g., \cite{Capitani:1994qn,Alexandrou:2016ekb} for previous application to the EMT operators and \cite{Capitani:2002mp} for a general setup). In this approach, an intermediate scheme between lattice and $\MSbar$ is not needed; the derivation of the renormalization factors can be obtained directly in the $\MSbar$ scheme by comparing the lattice bare Green's functions with the corresponding $\MSbar$-renormalized Green's functions calculated in DR. This approach can give reliable results only when higher-loop terms are negligible. The technical complexity of this approach effectively limits the applicability to one-loop order in most cases. A second approach regards the nonperturbative calculation of the mixing matrix by neglecting gauge-variant operators. These operators do not contribute to the calculation of physical quantities. However, they contribute to the correct extraction of operator renormalization factors from Green's functions with elementary external fields. This approach can give reliable results only when mixing effects by gauge-variant operators are small enough. A third approach is the combination of approaches 1 and 2 (e.g., \cite{Constantinou:2015ela,Alexandrou:2020sml}), where some elements of the mixing matrix are calculated nonperturbatively (e.g., the diagonal elements, or those related to lower-dimensional operators) while the remaining elements are calculated in perturbation theory. The mixing with gauge-variant operators is also omitted. 

\bigskip

In order to address the effects of gauge-variant operators, we propose an extension of the above approaches, including a semi-nonperturbative determination of the gauge-variant operators' contributions to the renormalization factors: The gluonic and fermionic parts of the gauge-variant operators can be calculated by lattice simulations, while the ghost part and/or the gauge-fixing terms can be obtained by lattice perturbation theory.       

\bigskip

Our proposed method can be applied in the present study of EMT operators and the nonperturbative calculation of their mixing matrix. The RI$'_1$ scheme, defined in Eqs. (\ref{RI1_1} -- \ref{RI1_5}), is not the optimal one, as it contains three operators with ghost and gauge-fixing terms and it entails the nonperturbative calculation of GFs with nonzero momentum transfer. Such calculation requires the use of two distinct momentum scales for the two external fields and the extrapolation of one momentum to zero, before calculating any renormalization factor. On the contrary, the RI$'_2$ scheme, defined in Eqs. (\ref{RI2_1} -- \ref{RI2_3}), is suitable for applying the proposed method. It entails calculating GFs of only three operators at zero momentum transfer. The first two operators $\mathcal{O}_1, \mathcal{O}_2$ are gauge invariant and, thus, their GFs are calculable by lattice simulations. The remaining operator $\mathcal{O}_6$ does not involve any gauge-fixing term; however, a ghost term is present. Writing, explicitly, $\mathcal{O}_6$,
\begin{eqnarray}
\mathcal{O}_{6 \mu \nu} \equiv \mathcal{O}_{4 \mu \nu} - \mathcal{O}_{5 \mu \nu} &=& \Big[ A_\mu^a \left( D_\rho F_{\rho \nu} \right)^a + A_\nu^a \left( D_\rho F_{\rho \mu} \right)^a - \frac{2}{d} \delta_{\mu \nu} A_\rho^a \left(D_\sigma F_{\sigma \rho}\right)^a \Big] \nonumber \\
&& - i g \Big[ A_\mu^a \bar{\psi} \gamma_\nu T^a \psi + A_\nu^a \bar{\psi} \gamma_\mu T^a \psi - \frac{2}{d} \delta_{\mu \nu} A_\rho^a \bar{\psi} \gamma_\rho T^a \psi \Big] \nonumber \\
&& + \Big[ \partial_\mu \bar{c}^a \partial_\nu c^a + \partial_\nu \bar{c}^a \partial_\mu c^a - \frac{2}{d} \delta_{\mu \nu} \partial_\rho \bar{c}^a \partial_\rho c^a \Big], 
\end{eqnarray}
(where $T^a$ are the generators of the $su(N_c)$ algebra), the first two terms can be investigated nonperturbatively by lattice simulations, while for the last term we content ourselves with its perturbative study. 

\bigskip

We note that the conditions of the RI$'_2$ scheme make use of amputated GFs. This may cause worry for the calculation of the gluonic GFs, where the inverse gluon propagator is needed in the process of the amputation; the lattice simulations commonly employ the Landau gauge, in which the gluon propagator is not invertible. However, setting to zero those components of the renormalization scale, which are parallel to the directions of the two external gluons, the amputation can be performed without inverting the whole gluon propagator. 

\bigskip

To explain in more detail the previous argument about the amputation of gluonic GFs in the Landau gauge, we consider the following amputated Green's function of the generic operator $\mathcal{O}_{\mu \nu}$:
\begin{equation}
{\langle A_{\rho} (q) \mathcal{O}_{\mu \nu} A_{\sigma} (-q) \rangle}_{\rm amp} = \sum_{\kappa, \lambda} {(D^{-1})}_{\rho \kappa} \langle A_\kappa (q) \mathcal{O}_{\mu \nu} A_\lambda (-q) \rangle {(D^{-1})}_{\lambda \sigma},
\label{Eq1}
\end{equation}  
where
\begin{equation}
D_{\rho \sigma} \equiv \langle A_\rho (q) A_\sigma (-q) \rangle = \frac{1}{q^2} \left(\delta_{\rho \sigma} - \frac{q_\rho q_\sigma}{q^2} \right) \Pi_T (q^2) + \alpha \frac{q_\rho q_\sigma}{{(q^2)}^2} \Pi_L (q^2) 
\end{equation}  
is the gluon propagator in a general gauge [$\Pi_T (q^2), \Pi_L (q^2)$ are scalar functions of $q^2$]. In the Landau gauge ($\alpha = 0$), the propagator is not invertible; however, if two components of the momentum q are zero, e.g., $q_1 = q_2 = 0$, then the first and second rows and columns of the propagator matrix take the values: $D_{1 \kappa} = D_{\kappa 1} = (1 / q^2) \delta_{\kappa 1} \Pi_T (q^2)$, $D_{2 \kappa} = D_{\kappa 2} = (1 / q^2) \delta_{\kappa 2} \Pi_T (q^2)$, respectively ($\kappa = 1, 2, 3, 4$). Of course, this is not true for the remaining rows and columns. Thus, the propagator takes a block-diagonal form  
\begin{equation} 
\begin{pmatrix}
\Pi_T (q^2)/q^2 & 0 & 0 & 0 \\
0 & \Pi_T (q^2)/q^2 & 0 & 0 \\
0 & 0 & D_{33} & D_{34} \\
0 & 0 & D_{43} & D_{44} 
\end{pmatrix}.
\end{equation}
The propagator is still not invertible. However, the upper block is invertible and can be inverted separately from the lower block. The latter can be inverted only in a general gauge $\alpha \neq 0$.

\bigskip

Now, going back to Eq. \eqref{Eq1} we observe that we do not need to calculate all the matrix elements of the inverse gluon propagator but only the $\rho$th row [for the calculation of ${(D^{-1})}_{\rho \kappa}, \ \kappa = 1, 2, 3, 4$] and the $\sigma$th column [for the calculation of ${(D^{-1})}_{\lambda \sigma}, \ \lambda = 1, 2, 3, 4$]. Thus, we do not need to invert the whole propagator matrix, but only the block containing $\rho, \sigma$ components if the propagator matrix is block diagonal. Choosing $q_{\rho} = q_{\sigma} = 0$, the propagator is indeed block diagonal, and thus, the amputation can be done successfully without inverting the whole gluon propagator. It follows that the Green's function $\langle A_{\rho} (q) O_{\mu \nu} A_{\sigma} (-q) \rangle |_{q = \bar{q}}$ in a momentum scale $\bar{q}$ with two vanishing components, e.g., $\bar{q}_1 = \bar{q}_2 = 0$, cannot be generally amputated in the Landau gauge; it can be amputated only in the case of ($\rho = 1$ or $\rho = 2$) and ($\sigma = 1$ or $\sigma = 2$). Similarly, $\langle A_{\rho} (q) O_{\mu \nu} A_{\sigma} (-q) \rangle |_{q = \bar{q}}$ with only one vanishing component, e.g., $\bar{q}_1 = 0$ can be amputated in the Landau gauge only when ($\rho = \sigma = 1$). Also, a ``democratic'' momentum renormalization scale cannot be applied in this case as the amputation cannot be implemented in the Landau gauge for this specific choice.

\bigskip

An alternative choice, ``RI$'_3$,'' is to consider nonamputated instead of amputated Green's functions,
\begin{eqnarray}
\widehat{G}_{gi} (q, -q) &\equiv & \langle A^a_\rho (q) {O_i}_{\mu \nu} (0) A^b_\sigma (-q) \rangle_{\rm nonamp}, \ (i = 1, 2, 6), \\
\widehat{G}_{qi} (q,q) &\equiv & \langle \psi (q) {O_i}_{\mu \nu} (0) \bar{\psi} (q) \rangle_{\rm nonamp}, \ (i = 1, 2, 6).
\end{eqnarray}
Then the conditions of Eqs. (\ref{RI2_1} -- \ref{RI2_3}) are replaced by
\begin{equation}
\frac{{\rm Tr} [\widehat{G}_{gi}^{{\rm RI}_3'} (q,-q)]}{N_c^2 - 1} \Bigg\vert_{\begin{smallmatrix}
\rho = \sigma, \\
\rho \neq (\mu, \nu), \\
q_\rho = 0, \\
q_\tau = \bar{q}_\tau, \ \forall \tau \neq \rho
\end{smallmatrix}} = \frac{{\rm Tr} [\widehat{G}_{gi}^{\rm tree} (q,-q)]}{N_c^2 - 1} \Bigg\vert_{\begin{smallmatrix}
\rho = \sigma, \\
\rho \neq (\mu, \nu), \\
q_\rho = 0, \\
q_\tau = \bar{q}_\tau, \ \forall \tau \neq \rho
\end{smallmatrix}} = \Bigg\{ \begin{matrix}
\ 2 \bar{q}_\mu \bar{q}_\nu / {(\bar{q}^2)}^2, \ &i=1 \\
\\
0, \ &i = 2, 6,
\end{matrix} \label{RI3_1}
\end{equation}
\begin{equation}
\frac{{\rm Tr} [\widehat{G}_{gi}^{{\rm RI}_3'} (q,-q)]}{N_c^2 - 1} \Bigg\vert_{\begin{smallmatrix}
\rho = \mu, \\
\sigma = \nu, \\
q_\rho = q_\sigma = 0, \\
q_\tau = \bar{q}_\tau, \ \forall \tau \neq (\rho, \sigma)
\end{smallmatrix}} = \frac{{\rm Tr} [\widehat{G}_{gi}^{\rm tree} (q,-q)]}{N_c^2 - 1} \Bigg\vert_{\begin{smallmatrix}
\rho = \mu, \\
\sigma = \nu, \\
q_\rho = q_\sigma = 0, \\
q_\tau = \bar{q}_\tau, \ \forall \tau \neq (\rho, \sigma)
\end{smallmatrix}} = \Bigg\{ \begin{matrix}
\ 1 / \bar{q}^2, \ &i=1 \\
0, \ &i = 2 \\
\ -2 / \bar{q}^2, \ &i=6,
\end{matrix} \label{RI3_2}
\end{equation}
\begin{equation}
\frac{1}{4 N_c} {\rm Tr} [\widehat{G}_{qi}^{{\rm RI}_3'} (q, q) \cdot \slashed{q}] \Bigg\vert_{\begin{smallmatrix}
q_\tau = \bar{q}_\tau, \forall \tau
\end{smallmatrix}} = \frac{1}{4 N_c} {\rm Tr} [\widehat{G}_{qi}^{\rm tree} (q,q) \cdot \slashed{q}] \Bigg\vert_{\begin{smallmatrix}
q_\tau = \bar{q}_\tau, \forall \tau
\end{smallmatrix}} = \Bigg\{ \begin{matrix}
0, \ &i = 1, 6 \\
\\
\ -i \bar{q}_\mu \bar{q}_\nu / \bar{q}^2, \ &i=2.
\end{matrix} \label{RI3_3}
\end{equation}
The second condition (Eq. \ref{RI3_2}) can be alternatively replaced by
\begin{equation}
\frac{{\rm Tr} [\widehat{G}_{gi}^{{\rm RI}_3'} (q,-q)]}{N_c^2 - 1} \Bigg\vert_{\begin{smallmatrix}
\rho \neq (\mu, \nu), \\
\sigma = \nu, \\
q_\sigma = 0, \\
q_\tau = \bar{q}_\tau, \ \forall \tau \neq \sigma
\end{smallmatrix}} = \frac{{\rm Tr} [\widehat{G}_{gi}^{\rm tree} (q,-q)]}{N_c^2 - 1} \Bigg\vert_{\begin{smallmatrix}
\rho \neq (\mu, \nu), \\
\sigma = \nu, \\
q_\sigma = 0, \\
q_\tau = \bar{q}_\tau, \ \forall \tau \neq \sigma
\end{smallmatrix}} = \Bigg\{ \begin{matrix}
\ -\bar{q}_\mu \bar{q}_\rho / {(\bar{q}^2)}^2, \ &i=1 \\
0, \ &i = 2 \\
\ (1 - \alpha_{{\rm RI}'}) \ \frac{\bar{q}_\mu \bar{q}_\rho}{{(\bar{q}^2)}^2}, \ &i=6,
\end{matrix} \label{RI3_2_alter}
\end{equation}
where the renormalization four-vector scale has one (instead of two) zero component. The third condition (Eq. \ref{RI3_3}) employing fermionic GFs could also involve amputated GFs, as they have no issues in the amputation process. The conversion factors from RI$'_3$ to the $\MSbar$ scheme coincide with those from RI$'_2$ to the $\MSbar$ scheme. For any other variant of RI$'$ scheme [e.g., Eq. \eqref{RI3_2_alter}], the conversion factors can be easily extracted from our expressions of the $\MSbar$-renormalized amputated GFs given in Eqs. (\ref{MSbar_first} -- \ref{MSbar_last}).  

\bigskip

Another possibility is to modify the RI$'_2$ renormalization scheme in a way that the sum $\mathcal{O}_1 + \mathcal{O}_2 + \mathcal{O}_6$ is a conserved quantity. In DR, the sum of the bare operators is conserved. However, this is not true on the lattice, where discretization effects violate translational invariance. A proper definition of the RI$'$ renormalization scheme can lead to a conserved sum of the renormalized operators even on the lattice. In the continuum, this is simple, as we explained in previous section; it requires the sum of RI$'$-renormalized operators to be equal to the sum of the bare operators. The corresponding lattice condition can be obtained by considering the WIs given in Eqs. (\ref{WI_cons3a}, \ref{WI_cons3b}). These WIs are extracted in DR; however, we can impose their validity also to the RI$'$-renormalized operators on the lattice. To avoid any issues regarding Landau-gauge fixing, these relations will give us three conditions by studying the specific choices of Lorentz and Dirac structures, obtained by the conditions of Eqs. (\ref{RI2_1} -- \ref{RI2_3}), i.e.,
\begin{equation}
\frac{{\rm Tr} [ \sum_{i=1,2,6} G_{gi}^{{{\rm RI}_2'}^{\rm cons}} (q,-q)]}{N_c^2 - 1} \Bigg\vert_{\begin{smallmatrix}
\rho = \sigma, \\
\rho \neq (\mu, \nu), \\
q_\rho = 0, \\
q_\tau = \bar{q}_\tau, \ \forall \tau \neq \rho
\end{smallmatrix}} = \frac{{\rm Tr} [\frac{1}{2} \left( q_\mu \frac{\partial}{\partial q_\nu} + q_\nu \frac{\partial}{\partial q_\mu} \right) {\left(D^{-1} (q) \right)}^{ab}_{\rho \rho}]}{N_c^2 - 1} \Bigg\vert_{\begin{smallmatrix}
\rho \neq (\mu, \nu), \\
q_\rho = 0, \\
q_\tau = \bar{q}_\tau, \ \forall \tau \neq \rho
\end{smallmatrix},} \label{RI2_1_cons}
\end{equation}
\begin{equation}
\frac{{\rm Tr} [\sum_{i=1,2,6} G_{gi}^{{{\rm RI}_2'}^{\rm cons}} (q,-q)]}{N_c^2 - 1} \Bigg\vert_{\begin{smallmatrix}
\rho = \mu, \\
\sigma = \nu, \\
q_\rho = q_\sigma = 0, \\
q_\tau = \bar{q}_\tau, \ \forall \tau \neq (\rho, \sigma)
\end{smallmatrix}} = \frac{{\rm Tr} [- \frac{1}{2} \left({\left(D^{-1} (q) \right)}^{ab}_{\rho \rho} + {\left(D^{-1} (q) \right)}^{ab}_{\sigma \sigma} \right)}{N_c^2 - 1} \Bigg\vert_{\begin{smallmatrix}
q_\rho = q_\sigma = 0, \\
q_\tau = \bar{q}_\tau, \ \forall \tau \neq (\rho, \sigma)
\end{smallmatrix},} \label{RI2_2_cons}
\end{equation}
\begin{equation}
\frac{1}{4 N_c} {\rm Tr} [\sum_{i=1,2,6} G_{qi}^{{{\rm RI}_2'}^{\rm cons}} (q,-q) \cdot \slashed{q}] \Bigg\vert_{\begin{smallmatrix}
q_\tau = \bar{q}_\tau, \forall \tau
\end{smallmatrix}} = \frac{1}{4 N_c} {\rm Tr} [\slashed{q} \cdot \frac{1}{2} \left( q_\mu \frac{\partial}{\partial q_\nu} + q_\nu \frac{\partial}{\partial q_\mu} \right) S^{-1} (q)] \Bigg\vert_{\begin{smallmatrix}
q_\tau = \bar{q}_\tau, \forall \tau
\end{smallmatrix}.} \label{RI2_3_cons}
\end{equation} 
In these conditions, the nonperturbative calculation of the discretized derivatives of gluon and quark propagators with respect to external momentum is needed. As we insert three new conditions, we must exclude three conditions from the previous definition of ${\rm RI}'_2$ scheme. For example, we exclude the operator $O_6$ from each condition [Eqs. (\ref{RI2_1} -- \ref{RI2_3})]. In this version of RI$'$, an operator with ghost fields is still involved and thus, a combination of perturbative and nonperturbative results is also needed.    

\bigskip

The proposed approach does not completely overcome the mixing effects stemming from gauge-variant operators. There are, in the literature, alternative methods for addressing this mixing. One method entails nonperturbative studies of BRST transformations and GFs with ghost fields implemented in the lattice simulations (see Refs. \cite{Ghiotti:2007qn,De:2019hov} and references therein). Another method investigates the nonperturbative renormalization of EMT on the lattice in a gauge-invariant way (Ref.~\cite{DallaBrida:2020gux}); in this method, WIs stemming from the conserved properties of the EMT are used in the framework of thermal QCD with a nonzero imaginary chemical potential. Finally, a gauge-invariant renormalization scheme, such as the X-space scheme \cite{Gimenez:2004me}, which considers gauge-invariant GFs in coordinate space, can be applied without the need of involving any gauge-variant operator. This scheme has not been applied before in the calculation of the mixing matrix of EMT operators. At the perturbative level, there is a work in progress by our group \cite{Costa:2020} in this direction. In  this case, the gauge-invariant GFs are constructed using only the gluon and quark EMT operators $\mathcal{O}_1$ and $\mathcal{O}_2$. Complications arise in this method. In order to calculate the $2 \times 2$ mixing matrix for the renormalization of $\mathcal{O}_1$ and $\mathcal{O}_2$, we need a total of four conditions. Three conditions can be obtained by studying two-point GFs between the two mixing operators (between themselves and between each other). However, a complete solution needs a fourth condition which cannot be obtained by any other two-point function. More details can be found in our forthcoming paper \cite{Costa:2020}.
    
\section{Summary}
\label{Summary}

In this paper, we study the two-loop renormalization and mixing of the gluon and quark EMT operators in dimensional regularization. To this end, we compute a set of two-point Green's functions, renormalized in $\MSbar$; from our results, one may directly deduce the conversion factors between $\MSbar$ and a large variety of RI$'$-like schemes which are appropriate for a nonperturbative extraction of renormalization functions through lattice simulations. We provide the conversion factors relating a number of specific versions of the RI$'$ scheme to $\MSbar$. 

\bigskip

We discuss in detail the application of our proposed schemes on the lattice and the construction of a nonperturbative renormalization program for the elimination of the operator-mixing effects. In particular, we propose a semi-nonperturbative approach, where perturbative and nonperturbative results are combined; the gluonic and fermionic contributions of gauge-variant operators, which mix with the gauge-invariant EMT operators, can be calculated nonperturbatively, while contributions from the ghost parts can be evaluated by lattice perturbation theory. Also, a different version of RI$'$ scheme is proposed, which leads to the determination of a conserved RI$'$-renormalized EMT on the lattice. Our approach, along with the results produced in this paper, can be applied in lattice simulations with the expectation of giving more reliable estimates. A complete elimination of mixing effects is currently under investigation by our group using the X-space renormalization scheme.   

\vspace*{1cm} 
\centerline{{\bf\large{{\bf{Acknowledgements}}}}}
\bigskip
H.P. acknowledges financial support from the project EXCELLENCE/0918/0066, funded by the Cyprus Research and Innovation Foundation. G.S. acknowledges financial support by the University of Cyprus, under the research programs entitled ``\textit{Quantum fields on the lattice}'' and ``\textit{Nucleon parton distribution functions using Lattice Quantum Chromodynamics}''.

\bibliographystyle{elsarticle-num}                     
\bibliography{EMT_References}

\end{document}